\documentclass[aps,pra,reprint,10pt,onecolumn,nofootinbib,longbibliography]{revtex4-1}

\usepackage{amsmath,amsthm,amsfonts,amssymb,bm}
\usepackage{mdframed,graphicx,array,MnSymbol,caption,subcaption,float,placeins}
\usepackage[usenames,dvipsnames]{xcolor}
\definecolor{OceanBlue}{rgb}{0,0.35,0.7} 
\usepackage{hyperref}
\hypersetup{colorlinks=true,allcolors=OceanBlue}

\newtheorem{thm}{Theorem}[section] % Theorem
\theoremstyle{definition}
\newtheorem{defn}{Definition}[section] % Definition
\newtheorem{remark}{Remark}[section] % Question

\newcommand{\pars}[1]{\left(#1\right)} % Parentheses
\newcommand{\bracs}[1]{\left[#1\right]} % Brackets
\newcommand{\lp}{\left(}  % Left paranthesis
\newcommand{\rp}{\right)} % Right paranthesis
\newcommand{\mr}{\mathrm} % Text in math mode
\newcommand{\mb}{\mathbf} % Bold math
\newcommand {\cl}{\mathcal} % Calligraphic math
\newcommand {\tsf} [1]{\textsf{#1}} % Upright sans serif math
\newcommand{\vectsym}{\boldsymbol}  %Vectors that are symbols
\DeclareMathOperator{\Tr}{Tr\,}            % operator trace
\DeclareMathOperator{\tr}{tr\,}            % trace
\newcommand*\diff{\mathop{}\!\mathrm{d}}
\DeclareMathOperator{\trans}{\textsf{T}} % Transpose
\DeclareMathOperator{\sech}{\mathrm{sech}} % Hyperbolic secant

\newcommand{\balpha}{\vectsym{\alpha}}  % Bold alpha
\newcommand{\bbeta}{\vectsym{\beta}} % Bold beta
\newcommand{\bphi}{\vectsym{\phi}} % Bold phi
\newcommand{\btheta}{\vectsym{\theta}} % Bold theta
\newcommand{\br}{\mb{r}} % Bold r

\newcommand {\ket}[1] {\left|{#1}\right\rangle}
\newcommand {\bra}[1] {\langle{#1}|}
\newcommand{\braket}[2]{\langle{#1}|{#2}\rangle}

\newcommand {\kets} [1] {\left|#1\right\rangle_{S}}
\newcommand {\bras} [1] {\left\langle#1\right|\hspace{0mm}_{S}}

\newcommand {\ketr} [1] {\left|#1\right\rangle_{R}}
\newcommand {\brar} [1] {\left\langle#1\right|\hspace{0mm}_{R}}

\newcommand {\keta} [1] {\left|#1\right\rangle_{A}}
\newcommand {\braa} [1] {\left\langle#1\right|\hspace{0mm}_{A}}

\newcommand {\ketas} [1] {\left|#1\right\rangle_{AS}}
\newcommand {\braas} [1] {\left\langle#1\right|\hspace{0mm}_{AS}}

\newcommand {\ketras} [1] {\left|#1\right\rangle_{RAS}}
\newcommand {\braras} [1] {\left\langle#1\right|\hspace{0mm}_{RAS}}

\newcommand{\varket}[1] {\left|{#1}\right\rrangle}
\newcommand{\varbra}[1] {\left\llangle #1\right|}
\newcommand{\varbraket}[2] {\left\llangle #1| #2 \right\rrangle}
\newcommand{\id}{\mathrm{id}}

\newcommand{\abs}[1]{\left | #1 \right |}   % Absolute value
\begin{document}
\title{Quantum sensing of phase-covariant optical channels}
\author{Ranjith Nair$^{1,}$}
\email{ranjith.nair@ntu.edu.sg}
\author{Mile Gu$^{1,2,}$}
\email{gumile@ntu.edu.sg}
\affiliation{$^1$Nanyang Quantum Hub, School of Physical and Mathematical Sciences,  \\
 Nanyang Technological University, 21 Nanyang Link, Singapore 639673 \\
$^2$Centre for Quantum Technologies, National University of Singapore, 3 Science Drive 2, Singapore 117543}

\date{\today}
\begin{abstract} 
We obtain universal (i.e., probe and measurement-independent) performance bounds on ancilla-assisted quantum sensing of multiple parameters of phase-covariant optical channels under energy and mode-number constraints. We first show that for any such constrained problem, an optimal ancilla-entangled probe can always be found whose reduced state on the modes probing the channel is diagonal in the photon-number basis. For parameters that are encoded in single-mode Gaussian channels, we derive a universal upper bound on the quantum Fisher information matrix that delineates the roles played by the energy and mode constraints. We illustrate our results for sensing of the transmittance of a thermal loss channel under both the no-passive-signature and passive-signature paradigms, and in the problem of sensing the noise variance of an additive-noise channel. In both cases, we show that two-mode squeezed vacuum probes are near-optimal under the constraints in the regime of low signal brightness, i.e., per-mode average photon number. More generally, our work sets down a uniform framework for readily evaluating universal limits for any sensing problem involving Gaussian channels.
\end{abstract}
\maketitle
Obtaining quantum-enabled enhancements in the sensing of parameters of optical -- and more generally, electromagnetic -- systems and channels is one of the key thrust areas of quantum technologies. Optical quantum metrology, progressing from its origins in precision phase sensing in interferometers using quantum states of light \cite{Cav81}, has  inspired the discovery of new measurement techniques that  improve the resolution of sensing parameters encoded in `classical' light as well\footnote{Here, `classical light' refers to light that is in a  probabilistic mixture of coherent states.} \cite{TNL16,Tsa19}. 

A third area in which quantum metrology has challenged our intuitions and led to new possibilities is that of \emph{quantum illumination} \cite{Llo08,TEG+08,Sha20} and related problems. Since quantum features of light are easily lost through decoherence, it is surprising that detecting a distant target buried in bright thermal noise using two-mode squeezed vacuum states of light enjoys an error probability exponent  that is $6$ dB (a factor of $4$) greater than the best classical probe of the same energy \cite{TEG+08}.  In fact, this quantum advantage exists \emph{only} in a range of noise brightness where characteristic quantum features such as entanglement and nonclassicality are lost \cite{NG20}. As such, there is a great effort at present to develop quantum illumination at  microwave wavelengths, where the noise naturally satisfies the high-brightness requirement \cite{Sha20,TB-WK21}. We direct the reader to refs.~\cite{PBG+18,BAB+18,PVS+20} for comprehensive reviews of the state of the art in optical quantum metrology.

In the present paper, we develop a general framework for predicting the presence and extent of quantum advantage from entanglement assistance for a wide class of  channels such as those encountered in quantum illumination, namely the \emph{phase-covariant} channels. Broadly speaking, these are channels whose action is agnostic to the phase of the input field. Inasmuch as time invariance is a common characteristic of physical interactions, such channels are ubiquitous. Allowing for arbitrary quantum probes that satisfy energy and mode-number constraints as well as arbitrary quantum measurements, we first show that the optimum probes for any such problem fall in a certain well-defined class.

Then, for the important subset of \emph{Gaussian} channels \cite{WPG+12,Ser17qcv,Hol19qsci}, we sharpen our result to get an explicit and easily evaluated upper bound on the quantum Fisher information matrix corresponding to \emph{any} given sensing problem. In sensing problems involving excess noise such as quantum illumination and gain sensing of linear amplifiers \cite{NTG22}, the number of modes available (e.g., the time-bandwidth product for temporal modes) figures as an important resource enhancing the performance \cite{TEG+08,NG20,NTG22,JDC22}.   The form of our upper bound  explicitly separates the contributions of  the energy and mode-number constraints, thus allowing easy comparison of these contributions. Finally, we illustrate the power of our bound by considering three specific sensing problems where it is approached by two-mode squeezed vacuum probes in the low-brightness regime.

The paper is organized as follows: In Sec.~\ref{sec:setup}, we introduce the general setup of quantum sensing and the relevant performance metrics and probe constraints. In Sec.~\ref{sec:pccsensing}, we define the phase-covariant channel sensing problem and prove Theorem~\ref{th:ndsoptimality} on the optimal form of the probes. In Sec.~\ref{sec:pcgcsensing}, we specialize to the case of phase-covariant Gaussian channels and give (in Theorem~\ref{th:qfimub})
 a universal upper bound on the performance of any probe satisfying the constraints. In Secs.~\ref{sec:tlsensing} and \ref{sec:anlsensing}, we illustrate the tightness of our general bound for the tasks of sensing the transmittance of a thermal loss channel and of sensing the noise variance of an additive-noise channel respectively. We conclude with a discussion of our results in Sec.~\ref{sec:discussion}.

\section{Quantum sensing Setup and Notation} \label{sec:setup}
\begin{figure}[tbp]
 \centering
      \includegraphics[trim=55mm 77mm 45mm 75mm, clip=true, scale=0.32]{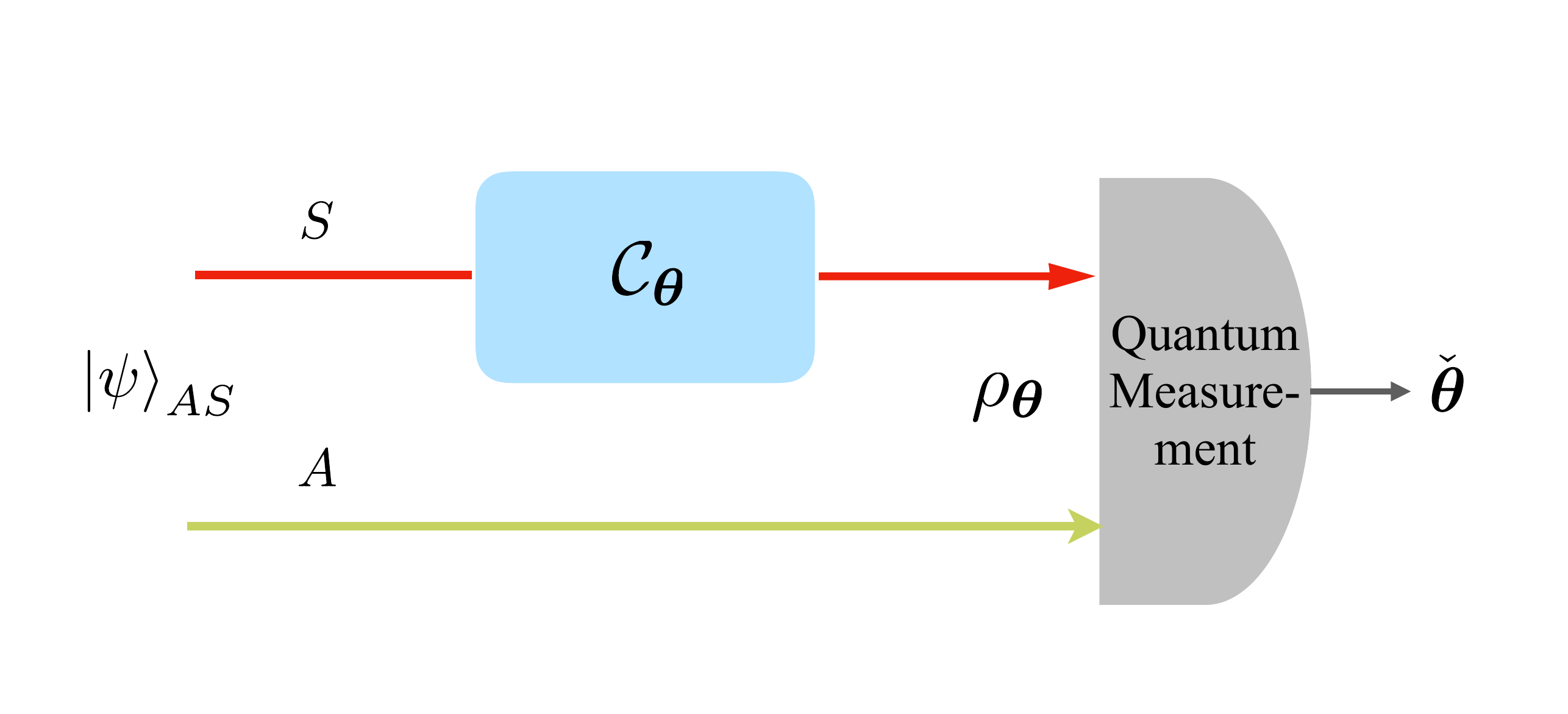}
%\onecolumngrid
\caption{A general ancilla-assisted parallel strategy for sensing a vector channel parameter $\btheta$: Each of $M$ signal ($S$) modes of a state $\ketas{\psi}$  entangled with an ancilla  system $A$ probes the unknown channel $\cl{C}_{\btheta}$. A subsequent joint measurement of the system and ancilla on the output state $\rho_{\btheta}$ (Eq.~\eqref{rhotheta}) generates an estimate $\check{\btheta}$ of $\btheta$.} \label{fig:channelsensing}
\end{figure}

Consider a family $\{\cl{C}_{\btheta}\}$ of quantum channels (completely positive trace-preserving maps) acting on an $M$-mode bosonic Hilbert space (called the \emph{signal} ($S$) system) and indexed by a vector of unknown parameters $\btheta = (\theta_1,\ldots, \theta_K) \in \boldsymbol{\Theta} \subset \mathbb{R}^K$. We consider the general ancilla-assisted parallel strategy illustrated in Fig.~\ref{fig:channelsensing} for estimating $\btheta$, where no restriction is placed on the  nature or dimensionality of the ancilla system $A$. The input to the channel (called the \emph{probe}) is a joint pure state $\ket{\psi}_{AS}$ of the signal and ancilla, which has the general form 
\begin{align} \label{probe}
\ketas{\psi} = \sum_{\mb{n}} \sqrt{p}_{\mb{n}}\ket{\chi_{\mb n}}_A \ket{\mb n}_S,
\end{align}
where $\ket{\mb n}_S = \ket{n_1}_{S_1}\ket{n_2}_{S_2}\cdots \ket{n_M}_{S_M}$ is an $M$-mode number state of $S$, $\{\ket{\chi_{\mb n}}_A\}$ are normalized (not necessarily orthogonal) states of $A$, and $p_{\mb n}$ is the probability mass function of $\mb{n}$. The resulting output state $\rho_{\btheta}$ is given by
\begin{align} \label{rhotheta}
\rho_{\btheta} = \pars{\mr{id}_A \otimes \cl{C}_{\btheta}}\Psi_{AS},
\end{align}
where $\Psi_{AS} = \ket{\psi}\braas{\psi}$ and $\mr{id}_A$ is the identity channel on $A$.

In the sequel, probes with the additional feature that $\braket{\chi_{\mb{n}}}{\chi_{\mb{n}'}}_{A} = \delta_{\mb{n},\mb{n}'}$ play a prominent role. This orthogonality of the $\left\{\keta{\chi_{\mb{n}}}\right\}$ implies that the reduced state of $S$ is diagonal in the number basis -- as such, we call such probes \emph{Number-Diagonal Signal (NDS)} probes.

To account for limitations on the energy that can be used to probe the channel, we constrain the average energy in the signal modes (that are assumed to be quasi-monochromatic) as  
\begin{align}
\bra{\psi}  \,\hat{I}_A \otimes \pars{\sum_{m=1}^M \hat{N}_m} \ketas{\psi} \equiv \bra{\psi} \,\hat{I}_A \otimes \hat{N}_S \ketas{\psi} =N,
\end{align}
 where $\hat{N}_m = \hat{a}_m^\dag \hat{a}_m$ is the number operator of the $m$-th signal mode and $\hat{I}_A$ is the identity on the ancilla system. This constraint can be simplified as 
\begin{align} \label{ec}
\sum_{n=0}^{\infty} n\,p_n = N,\;\;{\rm where}\;\; p_n = \sum_{\mb{n}\, :\, n_1 +\ldots + n_M = n} p_{\mb n}
\end{align}
is the probability mass function of the \emph{total} photon number in the signal modes. Note that a mixed-state probe can be purified using an additional ancilla resulting in a purification that is again of the form ~\eqref{probe} with the same $N$ and $M$. Thus, optimization over probes of the form of Eq.~\eqref{probe} suffices.

On the other hand, an arbitrary $M$-signal-mode \emph{classical} probe has the general form
\begin{align}\label{classicalprobe}
\rho_{AS} = \int_{\mathbb{C}^{M'}} \diff^{2M'} \balpha \int_{\mathbb{C}^{M}} \diff^{2M} \bbeta \,P(\balpha,\bbeta) \ket{\balpha}\bra{\balpha}_A\otimes\ket{\bbeta} \bra{\bbeta}_S,
\end{align}
where $\balpha = \pars{\alpha^{(1)}, \ldots, \alpha^{(M')}} \in \mathbb{C}^{M'}$ indexes $M'$-mode coherent states $\keta{\balpha}$ of $A$, $\bbeta = \pars{\beta^{(1)}, \ldots, \beta^{(M)}}\in \mathbb{C}^{M}$ indexes $M$-mode coherent states $\kets{\bbeta}$ of $S$, and $P\pars{\balpha,\bbeta} \geqslant 0$ is a probability distribution that allows arbitrary classical correlations between $S$ and $A$. The signal energy constraint now reads
%\begin{align}\label{cec}
$\int_{\mathbb{C}^{M'}} \diff^{2M'} \balpha \int_{\mathbb{C}^M} \diff^{2M} \bbeta\, P(\balpha,\bbeta) \pars{\sum_{m=1}^M \abs{\bbeta^{(m)}}^2} ={N}.$
%\end{align}

The number $M$ of available signal modes of both quantum and classical probes depends on operational constraints. For example, in the case of temporal modes, it equals the available time-bandwidth product. As such, it is an important resource in its own right and will be assumed to be given as well.

To quantify the estimation performance, we use the theory of quantum metrology \cite{Hel76,Par09,LYL+20}, which we briefly review.  For a given probe, the state family $\{\rho_{\btheta} = \pars{{\rm id}_A \otimes \cl{C}_{\btheta}}\pars{\ket{\psi}\bra{\psi}}\}$ gives rise to the $K \times K$  \emph{quantum Fisher information matrix (QFIM)} $\cl{K}_{\btheta}$ in the following way.  For each parameter $\theta_i$, there exists a Hermitian operator $\hat{L}_i$ (that depends on $\btheta$ in general) called the symmetric logarithmic derivative (SLD) operator satisfying $\partial_i \rho_{\btheta} \equiv \partial \rho_{\btheta}/ \partial \theta_i = \pars{\rho_{\btheta} \hat{L}_i + \hat{L}_i \rho_{\btheta}}/2$. The quantum Fisher information matrix (QFIM) $\cl{K}_{\btheta}$  is the $K \times K$ matrix whose $ij$-th entry is given by
$\bracs{\cl{K}_{\btheta}}_{ij} =  \Tr \rho_{\btheta} \pars{\hat{L}_i\hat{L}_j + \hat{L}_j\hat{L}_i}/2.$

The operational significance of the QFIM is as follows. Any measurement applied to the output  results in an \emph{estimate} vector $\check{\btheta} = \left (\check{\theta}_1,\ldots, \check{\theta}_K \right)$. The error covariance matrix $\Sigma$ of the estimate has the matrix elements $\Sigma_{ij} = \mathbb{E} \bracs{\pars{\check{\phi}_i - \phi_i}\pars{\check{\phi}_j - \phi_j}}$, where $\mathbb{E}$ denotes statistical expectation over the measurement results. If the estimate is  \emph{unbiased}, i.e., if $\mathbb{E}\bracs{\check{\btheta}} = \btheta$ for all $\btheta$,  we have the \emph{quantum Cram\'er-Rao bound} (QCRB): 
\begin{align} \label{qcrb}
\Sigma \geq \cl{K}_{\btheta}^{-1},
\end{align}
which is valid for \emph{any} (unbiased) quantum measurement. Here, the matrix inequality $A \geq B$ means that $A-B$ is positive semidefinite.

Our main concern in this paper is to obtain upper bounds (in this matrix-inequality sense) on the QFIM $\cl{K}_{\btheta}$ for the sensing of a wide and physically important class of channels. These  in turn lead to lower bounds on the error covariance matrix $\Sigma$ via Eq.~\eqref{qcrb}.  For the sensing of phase-covariant channels, the bounds obtained are universal in the sense of being valid for all probes $\ketas{\psi}$ satisfying the total signal energy and mode constraints. Since the resulting QCRBs are also intrinsically optimized over all possible measurement schemes, the resulting bounds constitute fundamental limits on the sensing performance.

\section{Sensing Phase-covariant channels} \label{sec:pccsensing}

We now define the class of phase-covariant channels which is the subject of this work.

\begin{defn}{[Phase-covariant channel]} \label{def:pcchannel}
A channel $\cl{C}$ acting on an $M$-mode bosonic Hilbert space $\cl{H}$  is called (jointly) \emph{phase-covariant} if
\begin{align}
\hat{U}\pars{\bphi} \cl{C} \pars{\rho} \hat{U}^{\dag}\pars{\bphi} &= \cl{C} \pars{\rho}
\end{align}
for all $\rho \in \cl{S}\pars{\cl{H}}$  and all $\bphi$, where $\bphi= (\phi_1,\cdots,\phi_M) \in [0,2\pi)^M$ is a vector of phase shifts and
\begin{align} \label{Uphi}
\hat{U}\pars{\bphi} & := \otimes_{m=1}^M  e^{-i\phi_m \hat{N}_m}
\end{align}
is the associated $M$-mode phase-shift unitary operator ($\hat{N}_m$ is the number operator of the $m$-th mode).
\end{defn}

In other words, the action of the channel commutes with arbitrary $M$-mode phase shifts. We note that phase-covariant channels are sometimes called \emph{gauge-covariant} channels in the literature -- see, e.g., \cite{Hol19qsci,DPTG16}. 

Determining whether or not a given channel is phase-covariant may be done in many ways. Such verification is very direct when the channel transformation is known in terms of the induced transformation on the characteristic function -- e.g., the Wigner characteristic function -- from input to output\footnote{Recall that for a mode with annihilation operator $\hat{a}$, the unitary operator       $\hat{D}(\xi) = e^{\xi\hat{a}^\dag - \xi^* \hat{a}}$ describes a displacement  in phase space by $\xi \in \mathbb{C}$. The (Wigner/Weyl) characteristic function of a state $\rho$ is defined as $\chi_{\rho}\pars{\xi}:= \Tr \rho \hat{D}\pars{\xi}$ and uniquely determines the state \cite{Hol11,Ser17qcv}. As such, a channel may be described completely by the transformation it effects on the input characteristic function. Extension to multimode systems is straightforward.}. 
In this way, it is easy to verify that common Gaussian channels such as the unitary phase-shift channels, the thermal attenuator and amplifier channels, and the additive `classical-noise' channel that adds circularly-symmetric noise in phase space \cite{Hol19qsci,Ser17qcv} are all phase-covariant. Beyond Gaussian channels, this class also includes generalized loss (resp.~amplifier) channels realized  by two-mode mixing (resp.~two-mode squeezing) of an input mode with a second `environment' mode in a number-diagonal initial state that is traced out at the output.  If the explicit state transformation is not available but the state evolution is given in the form of a Lindblad-form master equation, it is often still possible to verify phase covariance. This is the case for several multi-photon absorption and emission processes -- see, e.g., \cite{CHN+20}. 

Non-examples among Gaussian channels include the displacement channel, the (single-mode) squeezing channels, additive-noise channels introducing non-isotropic noise and the phase-conjugating attenuators and amplifiers \cite{Hol19qsci}  -- intuitively, such channels are not phase-covariant because they have `preferred' directions of action in phase space.

\begin{defn}{[Phase-covariant channel family]} \label{def:pcchannelfam}
A family of channels $\left\{\cl{C}_{\btheta}\right\}$ indexed by a vector parameter $\btheta = \pars{\theta_1, \ldots, \theta_K}^{\trans} \in \Theta \subset \mathbb{R}^K$ and acting on  an $M$-mode Hilbert space $\cl{H}$  is called a \emph{phase-covariant channel family} if  $\cl{C}_{\btheta}$ is  phase-covariant for all $\btheta \in \vectsym{\Theta}$.
\end{defn}

Our first main result is a general optimality claim for the class of NDS probes defined previously. Such probes have been previously studied in the context of many specific sensing and channel discrimination problems, e.g., the discrimination of beam-splitter channels \cite{Nai11}, phase estimation in the presence of loss \cite{KD-D10,Nai18}, and the joint estimation of phase and loss \cite{CDB+14}, to mention a few. Their optimality for discriminating or sensing arrays of beam-splitter channels in a vacuum environment was established in Ref.~\cite{NY11}. Their connection with phase covariance was elucidated in Ref.~\cite{SWA+18}, where they were shown to optimize a wide class of channel divergences between pairs of phase-covariant channels (Cf. Sec.~12 therein). The following theorem is an estimation-theoretic statement of their optimality for sensing phase-covariant channels.

\begin{thm}{[NDS probe optimality]}\label{th:ndsoptimality}
Given any family $\{\cl{C}_{\btheta}\}$ of phase-covariant channels acting on the $M$-mode signal Hilbert space $\cl{H}_S$ and parametrized by $\btheta \in \vectsym{\Theta}$. Among all probe states of the form 
\begin{align} \label{genprobe}
\ket{\psi} = \sum_{\mb{n}} \sqrt{p}_{\mb{n}}\ket{\chi_{\mb n}}_A \ket{\mb n}_S,
\end{align}
 with a given signal photon number distribution $\{p_{\mb{n}}\}$, the NDS probes 
\begin{equation}
\begin{aligned}  \label{ndsstate}
&\ket{\widetilde{\psi}} = \sum_{\mb{n}} \sqrt{p}_{\mb{n}}\ket{\widetilde{\chi}_{\mb n}}_A \ket{\mb n}_S \mbox{    with} \\
&\braket{\widetilde{\chi}_{\mb{n}}}{\widetilde{\chi}_{\mb{n}'}} = \delta_{\mb{n},\mb{n}'}
\end{aligned}
\end{equation} 
 maximize the QFIM $\cl{K}_{\btheta}$ on $\btheta$. In particular, an optimal probe state under an average signal energy constraint $N$ is also of the form of Eq.~\eqref{ndsstate}, for some $\{ p_{\mb{n}}\}$ satisfying the energy constraint.
 
\begin{proof} To begin with, assume that the probe \eqref{genprobe} has its signal photon number hard-limited to at most $N_0$ in each signal mode, i.e., that 
\begin{align} \label{truncatedprobe}
\ketas{\psi} = \sum_{\mb{0} \leq \mb{n} \leq \mb{N}_0} \sqrt{p}_{\mb{n}}\ket{\chi_{\mb n}}_A \ket{\mb n}_S,
\end{align}
where  $\mb{N}_0 = (N_0, \ldots, N_0)$ and  vector inequalities are understood to hold componentwise.
For $\br = \pars{r_1,\ldots, r_M}$ a vector of integers such that $\mb{0} \leq \br \leq \mb{N}_0 $, let $\br\cdot \hat{\mb{N}} := \sum_{m=1}^M r_m \hat{N}_m,$ and define the $M$-mode phase-shift unitary $\hat{U}(\br):= \exp\bracs{-i2\pi\pars{\br\cdot \hat{\mb{N}}}/(N_0+1)}$ and the phase-shifted probe state
\begin{align}
\ketas{\psi(\br)}:= \pars{\hat{I}_A \otimes\hat{U}(\br)}\ketas{\psi}
\end{align} 
Further, suppose that $R$ is a second ancilla (the `reference' system) of dimension  $(N_0+1)^M$ or greater, and let $\left\{\ketr{\br}: \mb{0} \leq \br \leq \mb{N}_0\right\}$ be an orthonormal state set thereof. Construct the augmented probe
\begin{align} \label{augprobe}
\ketras{\psi} := \frac{1}{\sqrt{(N_0+1)^M}} \sum_{\mb{0} \leq \br \leq \mb{N}_0} \ketr{\br}\ketas{\psi(\br)}.
\end{align}
We claim that the QFIM for $\btheta$ using the probe $\Psi_{RAS}:= \ket{\psi}\braras{\psi}$ satisfies $\cl{K}_{\btheta}\bracs{\Psi_{RAS}} \geq \cl{K}_{\btheta}\bracs{\Psi_{AS}}$, the QFIM for $\btheta$ using the original probe. To see this, note that given the probe $\Psi_{RAS}$, we can pass the signal system through the channel and then measure $R$ in the basis $\left\{\ketr{\br}\right\}$. Obtaining the outcome $\ketr{\br}$ (all $(N_0 + 1)^{M}$ outcomes are equally likely) collapses the state of the $AS$ system to $(\id_A \otimes \cl{C}_{\btheta})\pars{\ket{\psi(\br)}\braas{\psi(\br)}}$.  Applying $\hat{U}^\dag(\br)$ to $S$ then recovers the original modulated probe $(\id_A \otimes \cl{C}_{\btheta})\pars{\ket{\psi}\braas{\psi}}$ owing to the phase covariance of $\cl{C}_{\btheta}$. Thus, $\Psi_{RAS}$ has at least the same performance as $\Psi_{AS}$ (via the monotonicity of the QFIM \cite{Pet08qits}) and the assertion follows.

We now show that the reduced state in $S$ of the augmented probe $\Psi_{RAS}$ is diagonal in the multimode number basis. To see this, note that
 (all summations below are over vector indices ranging between $\mb{0}$ and $\mb{N}_0$)
 \begin{align}
 \Tr_{RA} \ket{\psi}\braras{\psi} &= \Tr_A \Tr_R \ket{\psi}\braras{\psi}\\
 &= (N_0+1)^{-M} \Tr_A \Tr_R \sum_{\br,\br'} \ket{\br}\brar{\br'}\otimes \ket{\psi(\br)}\braas{\psi(\br')} \\
 &= (N_0+1)^{-M}\Tr_A \sum_{\br} \ket{\psi(\br)}\braas{\psi(\br)},
 \end{align}
 since $\braket{\br}{\br'}_R = \delta_{\br,\br'}$. Further,
 \begin{align} 
\Tr_{RA} \ket{\psi}\braras{\psi} &= (N_0+1)^{-M}\Tr_A \sum_{\br} \ket{\psi(\br)}\braas{\psi(\br)}\\
&= (N_0+1)^{-M}\Tr_A \sum_{\mb{n},\mb{n}'} \sqrt{p_{\mb{n}}p_{\mb{n}'}}\ket{\chi_{\mb{n}}}\braa{\chi_{\mb{n}'}} \otimes \pars{\sum_{\br} \hat{U}(\br) \ket{\mb{n}}\bra{\mb{n}'}\hat{U}^\dag(\br)} \\
&=  (N_0+1)^{-M} \sum_{\mb{n},\mb{n}'} \sqrt{p_{\mb{n}}p_{\mb{n}'}} \braket{\chi_{\mb{n}'}} {\chi_{\mb{n}}}_A \pars{ \sum_{\br} e^{-i2\pi\br\cdot \pars{\mb{n} - \mb{n}'}/(N_0+1)}} \ket{\mb{n}}\bra{\mb{n}'} \\
&= \sum_{\mb{n}} p_{\mb{n}} \ket{\mb{n}}\bras{\mb{n}},
\end{align}
since the sum over $\br$ in the parantheses equals $(N_0 + 1)^M \delta_{\mb{n},\mb{n}'}$.

By the Schmidt decomposition there exist orthonormal states $\{\ket{\widetilde{\chi}_n}_{A'} \}$ of an ancilla system $A'$ and an NDS probe
\begin{align}
\ket{\widetilde{\psi}}_{A'S} = \sum_{n} \sqrt{p_n}\ket{\widetilde{\chi}_n}_{A'} \ket{ n}_S
\end{align}
with the same reduced state on $S$ as the augmented probe $\ketras{\psi}$. Such states are related by an isometry taking $AR$ to $A'$. Since such an isometry commutes with the action of the channel, $\cl{K}_{\btheta}\bracs{\widetilde{\Psi}_{A'S}} = \cl{K}_{\btheta}\bracs{\Psi_{RAS}}$. We have thus shown that, for any probe of the form \eqref{truncatedprobe} with photon number hard-limited to $N_0$, an NDS probe with the same signal photon number distribution has equal or greater QFIM in the matrix-inequality sense.

For a general probe of the form \eqref{genprobe}, consider the sequence of probes
\begin{align}
\ketas{\psi(N_0)} := \frac{\sum_{\mb{0} \leq \mb{n} \leq \mb{N}_0} \sqrt{p_{\mb{n}}} \keta{\chi_{\mb{n}}}\kets{\mb{n}}} {\sqrt{\sum_{\mb{0} \leq \mb{n} \leq \mb{N}_0} p_{\mb{n}}}}
\end{align}
with $N_0 = 0,1, \ldots$.
By the above argument, each such probe can --  without decreasing the QFIM -- be replaced by the NDS probe 
\begin{align}
\ketas{\widetilde{\psi}(N_0)} := \frac{\sum_{\mb{0} \leq \mb{n} \leq \mb{N}_0} \sqrt{p_{\mb{n}}} \keta{\widetilde{\chi}_{\mb{n}}}\kets{\mb{n}}} {\sqrt{\sum_{\mb{0} \leq \mb{n} \leq \mb{N}_0} p_{\mb{n}}}},
\end{align}
for a definite choice of the set of orthogonal states $\left\{\keta{\widetilde{\chi}_{\mb{n}}}\right\}_{\mb{0} \leq \mb{n} \leq \mb{N}_0}$ that remains fixed as $N_0$ is increased. 

 As $N_0 \rightarrow \infty$, $\ketas{\psi(N_0)} \rightarrow \ketas{\psi}$, so that by continuity the QFIM of the NDS probe
 \begin{align}
 \ketas{\widetilde{\psi}} = \sum_{\mb{n} \geq \mb{0}} \sqrt{p_{\mb{n}}} \keta{\widetilde{\chi}_{\mb{n}}}\kets{\mb{n}} = \lim_{N_0 \rightarrow \infty} \ketas{\widetilde{\psi}(N_0)} 
 \end{align}
 is greater than or equal to that of the probe \eqref{genprobe}, proving the theorem.
\end{proof}
\end{thm}

\section{Sensing phase-covariant Gaussian channels} \label{sec:pcgcsensing}

In the remainder of this paper, we consider the sensing of parameters of single-mode \emph{phase-covariant Gaussian channels} (PCGCs). After the unitary phase-shift channels, perhaps the most commonly arising PCGC is the \emph{quantum-limited attenuator} (or \emph{pure-loss}) channel \cite{Hol19qsci,Ser17qcv} of transmittance $\eta \leq 1$, denoted $\cl{L}_{\eta}$. The channel is defined by the following action on the  characteristic function $\chi_{\rm in}(\xi) := \Tr \rho \hat{D}(\xi)$ of the input state $\rho$:
\begin{align} \label{losstrans}
\chi_{\rm out}(\xi) = \chi_{\rm in}(\sqrt{\eta} \,\xi)\, e^{-(1-\eta)\abs{\xi}^2/2}. 
\end{align} 
Another  PCGC is the \emph{quantum-limited amplifier} channel \cite{Hol19qsci,Ser17qcv} of gain $G \geq 1$ (denoted $\cl{A}_{G}$), defined by the characteristic function transformation:
\begin{align} \label{gaintrans}
\chi_{\rm out}(\xi) = \chi_{\rm in}(\sqrt{G} \,\xi)\, e^{-(G-1)\abs{\xi}^2/2}. 
\end{align} 
A key classification theorem for single-mode PCGCs states that any such channel $\cl{C}$ can be written -- save for an additional phase shift -- as a composition 
\begin{align} \label{gencomp}
\cl{C} = \cl{A}_G \circ \cl{L}_{\eta}
\end{align}
 of a  quantum-limited loss channel followed by a quantum-limited amplifier with suitably chosen $\eta$ and $G$ \cite{CGH06,Hol19qsci}. 
 
 In earlier work, we studied the ancilla-assisted sensing of quantum-limited loss \cite{Nai18loss} and gain \cite{NTG22} channels using arbitrary multimode probes. It was shown that any NDS probe satisfying the energy and mode constraints is a quantum-optimal probe, which is consistent with  Theorem \ref{th:ndsoptimality}. The following result leverages this work on the quantum-limited channels to place an upper bound on the QFIM for sensing parameters of any PCGC family that admits a decomposition of the form of Eq.~\eqref{gencomp}.

\begin{thm}{[Upper bound on the QFIM for PCGCs]}\label{th:qfimub}
Let a parameter $\btheta = \pars{\theta_1,\ldots, \theta_K} \in \vectsym{\Theta} \subset \mathbb{R}^K$ with $K \leq 2$ be encoded in a family of single-mode phase-covariant Gaussian channels $\left\{\cl{C}_{\btheta}\right\}$ which admit the decomposition
\begin{align} \label{decomp}
\cl{C}_{\btheta} = \cl{A}_{G(\btheta)} \circ \cl{L}_{\eta\pars{\btheta}},\;\;\ \btheta \in \vectsym{\Theta},
\end{align}
as an attenuator-amplifier cascade with $\eta\pars{\btheta}$ and $G\pars{\btheta}$ being twice-differentiable functions. Let $\rho_{AS}$ be a possibly ancilla-entangled $M$-signal-mode probe satisfying the signal energy constraint $\Tr \rho_{AS} \hat{N}_S = N$. The quantum Fisher information matrix $\cl{K}_{\btheta}$ on  $\btheta$ of the state family 
\begin{align}
\rho_{\btheta} = \pars{\mr{id}_A \otimes \cl{C}_{\btheta}^{\otimes M}}\rho_{AS}
\end{align}
satisfies the matrix inequality $\cl{K}_{\btheta} \leq \cl{\widetilde{K}}_{\btheta}$ where the $ij$-th entry of $\cl{\widetilde{K}}_{\btheta}$ is given by
\begin{align} \label{QFIMub}
\bracs{\widetilde{\cl{K}}_{\btheta}}_{ij} &= \frac{N}{\eta(1-\eta)}\frac{\partial \eta}{\partial \theta_i}\frac{\partial \eta}{\partial \theta_j} + \frac{\eta N + M} {G(G-1)} \frac{\partial G}{\partial \theta_i}\frac{\partial G}{\partial \theta_j},
\end{align}
for $1\leq i,j \leq K$.
\begin{proof}
Any given probe $\rho_{AS}$ can first be purified using an additional ancilla system $A'$ without changing its statistics (and hence its average signal photon number) on $S$. We can therefore assume that the probe state is pure.
Accordingly, let $\ket{\psi}_{AS} = \sum_{\mb{n} \geq \mb{0}} \sqrt{p_{\mb{n}}} \keta{\chi_{\mb{n}}}\kets{\mb{n}}$ be any $M$-signal mode probe satisfying the energy constraint. In order to get an upper bound on the QFIM $\cl{K}_{\btheta}$, we can assume by Theorem \ref{th:ndsoptimality} that $\Psi_{AS}$ is NDS. Using the decomposition  in Eq.~\eqref{decomp}, we can write the output state 
\begin{align}
\rho_{\btheta} &= \pars{\mr{id}_A \otimes \cl{C}_{\btheta}^{\otimes M}} \Psi_{AS} \\
			&= \pars{\mr{id}_A \otimes \cl{A}_{G}^{\otimes M}}\pars{\mr{id}_A \otimes \cl{L}_{\eta}^{\otimes M}} \Psi_{AS} \\
			&= \pars{\mr{id}_A \otimes \cl{A}_{G}^{\otimes M}}\pars{\sum_{\mb{l} \geq \mb{0}} p_{\mb{l}}\pars{\eta} \Psi_{\mb{l}}\pars{\eta}} \\
			&=  \sum_{\mb{l} \geq \mb{0}} p_{\mb{l}}\pars{\eta} \pars{\mr{id}_A \otimes \cl{A}_{G}^{\otimes M}} \Psi_{\mb{l}}\pars{\eta}\\
			&\equiv  \sum_{\mb{l} \geq \mb{0}} p_{\mb{l}}\pars{\eta} \rho\pars{G,\eta,\mb{l}}. \label{statedecomp}
\end{align}
Here, $\mb{l} = (l_1, \ldots, l_M)$ can be interpreted as the number of photons in each signal mode lost to the environment of the attenuator channel in a hypothetical implementation  of $\cl{C}_{\btheta}$ via Eq.~\eqref{decomp} with
\begin{align} \label{pleta}
p_{\mb{l}}\pars{\eta} = \sum_{\mb{n} \geq \mb{l}} p_{\mb{n}} \pars{\prod_{m=1}^M \eta^{n_m -l_m} (1-\eta)^{l_m}}
\end{align}
and $\Psi_{\mb{l}}\pars{\eta} = \ket{\psi_{\mb{l}}\pars{\eta}}\braas{\psi_{\mb{l}}\pars{\eta}}$ being the NDS state
\begin{align} \label{psileta}
\ketas{\psi_{\mb{l}}\pars{\eta}} &= \sum_{\mb{k} \geq \mb{0}} \sqrt{p_{\mb{k}|\mb{l}}\pars{\eta}} \keta{\chi_{\mb{k} + \mb{l}}}\kets{\mb{k}},\mbox{ where}\\
 p_{\mb{k}|\mb{l}}\pars{\eta} &= p_{\mb{k}+\mb{l}}\pars{\prod_{m=1}^M \eta^{k_m} (1-\eta)^{l_m}}/p_{\mb{l}}(\eta)
\end{align}
is a conditional probability distribution.
Derivations of the above expressions can be found in Ref.~\cite{Nai18loss}. The key point here, however, is that Eq.~\eqref{statedecomp} provides a decomposition of $\rho_{\btheta}$ in terms of the conditional states $\left\{\rho\pars{G,\eta,\mb{l}}\right\}$ occurring with the parameter-dependent probabilities $\left\{p_{\mb{l}}\pars{\eta}\right\}$. As such, we can apply the extended convexity of the QFIM \cite{NAW+16,AR15} to write the matrix inequality
\begin{align} \label{extconvexity}
\cl{K}_{\btheta}\bracs{\rho_{\btheta}} &\leq \cl{J}_{\btheta}\bracs{\mb{L}} + \sum
_{\mb{l} \geq \mb{0}} p_{\mb{l}}\pars{\eta} \cl{K}_{\btheta}\bracs{\rho\pars{G,\eta,\mb{l}}}.
\end{align}
Here, $\cl{J}_{\btheta}\bracs{\mb{L}}$ is the (classical) Fisher information matrix (FIM) on $\btheta$ corresponding to the random variable $\mb{L}$ representing the loss pattern of photons to the environment of the attenuator (which is distributed according to $\left\{p_{\mb{l}}\pars{\eta}\right\}$) and $\cl{K}_{\btheta}\bracs{\rho\pars{G,\eta,\mb{l}}}$ is the QFIM of the conditional state $\rho\pars{G,\eta,\mb{l}}$.

We now compute a generic  term $\cl{K}_{\btheta}\bracs{\rho\pars{G,\eta,\mb{l}}}$ of the sum in Eq.~\eqref{extconvexity}, where
\begin{align} \label{conditionaloutputs}
\rho\pars{G,\eta,\mb{l}} = \pars{\mr{id}_A \otimes \cl{A}_{G}^{\otimes M}} \Psi_{\mb{l}}\pars{\eta}.
\end{align}
To do so, we first compute the fidelity (The fidelity between states $\rho$ and $\rho'$ is $F\pars{\rho,\rho'} := \Tr \sqrt{\sqrt{\rho} \,\rho' \sqrt{\rho}}$ ) between two instances of the above state with parameters $(\eta, G)$ and $(\eta',G')$ respectively. This can be done by modifying a similar calculation from Ref.~\cite{NTG22} (See Appendix~\ref{app:appendix}) to get
\begin{align} \label{conditionalfid}
F\pars{\rho\pars{G,\eta,\mb{l}},\rho\pars{G',\eta',\mb{l}}} = \sum_{\mb{k} \geq \mb{0}} \sqrt{ p_{\mb{k}|\mb{l}}\pars{\eta} p_{\mb{k}|\mb{l}}\pars{\eta'}}\; \nu^{k+M},
\end{align}
where $k = \sum_{m=1}^M k_m$ and $\nu = \pars{\sqrt{GG'} - \sqrt{\pars{G-1}\pars{G'-1}}}^{-1}$. The $ij$-th entry of the QFIM is then given by the formula \cite{LYL+20}:
\begin{align} \label{condqfimelement}
\cl{K}_{\btheta}\bracs{\rho\pars{G,\eta,\mb{l}}}_{ij} = -4\pars{\partial'_i\partial'_j F\bracs{\rho\pars{G,\eta,\mb{l}},\rho\pars{G',\eta',\mb{l}}}}  \Big\vert_{\btheta'=\btheta},
\end{align}
where $\partial'_i := \partial/\partial \theta'_i$. In computing the above, the quantity
\begin{align} \label{condbhattacoeff}
B_{\mb{K}|\mb{l}}\pars{\eta,\eta'} := \sum_{\mb{k} \geq \mb{0}} \sqrt{p_{\mb{k}|\mb{l}}\pars{\eta}p_{\mb{k}|\mb{l}}\pars{\eta'}} 
\end{align} 
that arises is the classical fidelity (or Bhattacharyya coefficient \cite{Kai67,FvdG99}) between the distributions of the residual photon number pattern $\mb{K}$ in the $S$ modes after the input state passes through $\cl{L}_{\eta}^{\otimes M}$ and $\cl{L}_{\eta'}^{\otimes M}$ respectively, both conditioned on the loss pattern $\mb{l}$.
Evaluation of the derivatives appearing in Eq.~\eqref{condqfimelement} at $\btheta'=\btheta$ (i.e.,  setting $G'=G$ and $\eta'=\eta$) is aided by the observation that 
\begin{align}
\partial'_i B_{\mb{K}|\mb{l}}\pars{\eta,\eta'} \big\vert_{\eta'= \eta} &= \frac{\partial  B_{\mb{K}|\mb{l}}\pars{\eta,\eta'}}{\partial \eta'}\Bigg\vert_{\eta'= \eta} \frac{\partial \eta}{\partial \theta_i} = 0 \\
\partial'_i \nu \big \vert_{G'=G} &=\frac{\partial  \nu}{\partial G'}\Bigg\vert_{G'= G} \frac{\partial G}{\partial \theta_i} = 0
\end{align}
for $i=1,2$ --- this is because $B_{\mb{K}|\mb{l}}\pars{\eta,\eta'}$ and $\nu$ take their maximum values of unity for $\eta' \rightarrow \eta$ and $G' \rightarrow G$ respectively. The final result is
\begin{align} \label{condqfielementresult}
\cl{K}_{\btheta}\bracs{\rho\pars{G,\eta,\mb{l}}}_{ij} &= -4 \pars{\partial'_i \partial'_j B_{\mb{K}|\mb{l}}\pars{\eta,\eta'}}\bigg\vert_{\eta'=\eta} -4 \sum_{\mb{k} \geq \mb{0}} (k+M)\pars{\partial'_i \partial'_j \nu}\bigg\vert_{G'=G}.
\end{align}
Note that, owing to the connection of the Bhattacharyya coefficient and the classical FIM, the first term is simply the $ij$-th entry of the FIM $\cl{J}_{\btheta}\bracs{\mb{K}|\mb{l}}$ of $\mb{K}$ conditioned on the value $\mb{l}$. Putting everything together, the $ij$-th entry of the right-hand side of \eqref{extconvexity} becomes
\begin{align}
\bracs{\widetilde{\cl{K}}_{\btheta}}_{ij} &= \bracs{\cl{J}_{\btheta}\bracs{\mb{L}}}_{ij} + \sum_{\mb{l} \geq \mb{0}} p_{\mb{l}}\pars{\eta} \bracs{\cl{J}_{\btheta}[\mb{K}|\mb{l}]}_{ij} -4 \sum_{\mb{l} \geq \mb{0}} p_{\mb{l}}\pars{\eta} (k+M)\pars{\partial'_i \partial'_j \nu}\bigg\vert_{G'=G}\\
&= \bracs{\cl{J}_{\btheta}\bracs{\mb{L}} + \cl{J}_{\btheta}\bracs{\mb{K}|\mb{L}}}_{ij} - 4(\eta N + M) \pars{\partial'_i \partial'_j \nu}\bigg\vert_{G'=G}\\
&= \bracs{\cl{J}_{\btheta}\bracs{\mb{L},\mb{K}}}_{ij} - 4(\eta N + M) \pars{\partial'_i \partial'_j \nu}\bigg\vert_{G'=G}
\end{align}
where we have used the chain rule for the classical FIM $\cl{J}_{\btheta}\bracs{\mb{L},\mb{K}}$ of the combined observation $\pars{\mb{L},\mb{K}}$ \cite{Zam98} and observed that the unconditional average of the total photon number in the signal modes after the attenuator is simply $\eta N$. Since the statistics of $\pars{\mb{L},\mb{K}}$ depend on $\btheta'$ only via $\eta'$ and $\nu$ depends on $\btheta'$ only via $G'$, we have
\begin{align}
\bracs{\widetilde{\cl{K}}_{\btheta}}_{ij} &= \cl{J}_{\eta}\bracs{\mb{L},\mb{K}}\frac{\partial \eta}{\partial \theta_i}\frac{\partial \eta}{\partial \theta_j} + \frac{\eta N + M} {G(G-1)} \frac{\partial G}{\partial \theta_i}\frac{\partial G}{\partial \theta_j},
\end{align}
where we have used $-4\frac{\partial^2 \nu}{\partial G'^2}\Big\vert_{G'=G} = \frac{1}{G(G-1)}$ \cite{NTG22}. For the final step, we note that access to $\pars{\mb{L},\mb{K}}$ at the attenuator output is equivalent to also having access to the result of a measurement of the basis $\left\{\keta{\chi_{\mb{n}}}\right\}$ on $A$ since this result is necessarily $\keta{\chi_{\mb{K}+\mb{L}}}$. It was shown in \cite{Nai18loss} that for NDS probes this joint measurement attains the quantum-optimal value
\begin{align}
\cl{J}_{\eta} \bracs{\mb{K},\mb{L}} = \frac{N}{\eta(1-\eta)},
\end{align}
from which the result follows.
 \end{proof}
\end{thm}

Several remarks are in order at this point.

\remark{To avoid confusion, it should be mentioned that Theorem~\ref{th:qfimub} gives a valid bound  whether or not the channel $\cl{C}_{\btheta}$ is actually implemented as a cascade of attenuator and amplifier channels. On the other hand, if such an implementation is made and an observer is able to measure the loss pattern $\mb{L}$ at the output of the attenuator stage in addition to having access to $AS$ at the output of the amplifier stage, it appears that the inequality \eqref{extconvexity} used in the derivation becomes an equality and the bound is tight. Since such access is not available in general, the bound is not guaranteed to be tight even for NDS probes (though it is for the pure-loss and quantum-limited gain channels). We show in the sequel that it can still be closely approached in many problems of interest.
} \label{rem1}

\remark{If we take $K=1$ and set $\theta_1 = \eta$ and $G=1$, the sensing problem reduces to that of estimating the transmittance of a pure-loss channel and Eq.~\eqref{QFIMub} reproduces the known quantum limit \cite{Nai18loss}. Similarly, setting  $K = \eta=1$ and $\theta_1=G$ recovers the problem of sensing the gain of a quantum-limited amplifier, for which Eq.~\eqref{QFIMub} again reproduces the achievable quantum limit \cite{NTG22}. More generally, for a given sensing problem, the simple form of the bound in Eq.~\eqref{QFIMub} enables easy analytical computation once the functions $\eta(\btheta)$ and $G(\btheta)$ are at hand.
} \label{rem2}

\remark{From Eq.~\eqref{QFIMub}, we see that the generic $\cl{\widetilde{K}}\bracs{\btheta}$ splits into a sum of the form
\begin{align} \label{modalphot}
\cl{\widetilde{K}}_{\btheta} &= \cl{\widetilde{K}}^{\tsf{pp}}_{\btheta} N + \cl{\widetilde{K}}^{\tsf{pm}}_{\btheta} M
\end{align}
where $\cl{\widetilde{K}}^{\tsf{pp}}_{\btheta}$ and $\cl{\widetilde{K}}^{\tsf{pm}}_{\btheta}$ are respectively \emph{per-photon (pp)} and \emph{per-mode (pm)} contributions to the QFIM upper bound with the respective $ij$-th entries
\begin{align} \label{pppm}
\bracs{\cl{\widetilde{K}}^{\tsf{pp}}_{\btheta}}_{ij} &= \frac{1}{\eta(1-\eta)}\frac{\partial \eta}{\partial \theta_i}\frac{\partial \eta}{\partial \theta_j} + \frac{\eta } {G(G-1)} \frac{\partial G}{\partial \theta_i}\frac{\partial G}{\partial \theta_j},\\
\bracs{\cl{\widetilde{K}}^{\tsf{pm}}_{\btheta}}_{ij} &=  \frac{1 } {G(G-1)} \frac{\partial G}{\partial \theta_i}\frac{\partial G}{\partial \theta_j}.
\end{align} \label{rem3}
Thus, the generic PCGC sensing performance cannot exceed `standard quantum limit (SQL)' scaling in either of the resources $M$ or $N$.  The per-mode contribution is nonzero whenever the vacuum state has metrological power -- the expression above clarifies the situations for which this is the case.}

%\section{Applications} \label{sec:applications}

\section{Application: Sensing transmittance of a thermal loss channel} \label{sec:tlsensing}

\begin{figure}[tbp] 
 \centering
      \includegraphics[trim=25mm 80mm 17mm 65mm, clip=true, scale=0.5]{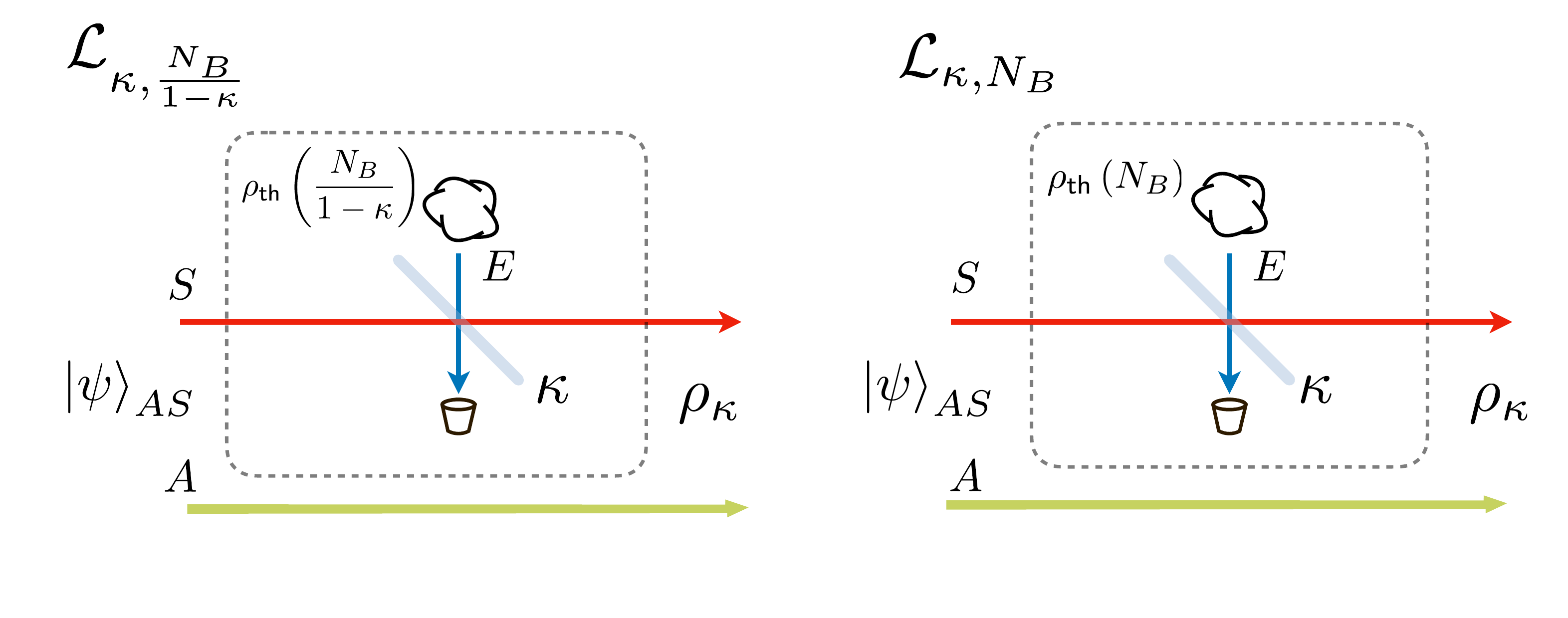}
%\onecolumngrid
\caption{Two paradigms for sensing the transmittance of a thermal loss channel: (Left) The No Passive Signature (NPS) paradigm in which the environment noise brightness is adjusted to leave the output state independent of $\kappa$ for a vacuum input. (Right) The Passive Signature (PS) paradigm in which the noise brightness remains constant at $N_B$ for any value of $\kappa$.} \label{fig:noisyatt}
\end{figure} 

As our first application, we consider the estimation of the transmittance of a thermal channel \cite{Hol19qsci,Ser17qcv}. This channel is physically realized as an interaction between the signal mode and an \emph{environment} ($E$) mode  in a thermal state via a beam splitter of transmittance $\kappa$ (See Fig.~\ref{fig:noisyatt}).  The brightness (average per-mode photon number) of the background $N_B>0$ is taken to be known a priori.  The channel models free-space and guided-wave  electromagnetic channels with background  thermal noise. While $N_B \sim 0$ may be achieved at optical frequencies in many situations, $N_B \gg 1$  at microwave frequencies at room temperature.
The quantum channel induced on $S$ when $E$ is traced out is denoted $\cl{L}_{\kappa,N_B}$. The case $N_B=0$ corresponds to the pure-loss channel $\cl{L}_{\kappa}$ introduced earlier. The characteristic function transformation effected by $\cl{L}_{\kappa,N_B}$ is
\begin{align} \label{noisyatt}
\chi_{\rm out}(\xi) = \chi_{\rm in}(\sqrt{\kappa} \,\xi)\, e^{-(1-\kappa)\pars{N_B + \frac{1}{2}}\abs{\xi}^2}. 
\end{align} 

Two paradigms for sensing the transmittance of noisy attenuators are prevalent in the literature. The first, which is mainly used in the literature of target detection via quantum illumination (QI) (see, e.g., ref.~\cite{Sha20} and references therein)  employs a `no passive signature'  \cite{TEG+08} or  `normalized noise' assumption \cite{JDC22} in which the environment brightness is adjusted as a function of $\kappa$. The second paradigm corresponds exactly to the transformation \eqref{noisyatt} above. We discuss our results in connection with both paradigms in turn.

\subsection{No Passive Signature Assumption} \label{sec:nps}

Under the \emph{No Passive Signature (NPS)} assumption, the environment is taken to be in a thermal state of brightness $N_B/(1-\kappa)$. As a result, the output state for a vacuum (hence `passive') probe is a thermal state $\rho_{\tsf{th}}\pars{N_B}$ of brightness $N_B$, which carries no dependence (`signature') on $\kappa$. The assumption can be justified for standoff detection and sensing scenarios for which $\kappa \ll 1$ and often affords ease of analysis. Transmittance sensing under the NPS assumption has been discussed in the works \cite{SLHG-R+17,NG20,JDC22,GRG+23}.

To describe the sensing problem in the framework of Sec.~\ref{sec:pcgcsensing}, we set $K=1$ and $\theta_1 = \kappa$ and $\cl{C}_{\kappa} = \cl{L}_{\kappa, N_B/(1-\kappa)}$ in Theorem~\ref{th:qfimub}. It is readily shown via Eqs.~\eqref{losstrans}-\eqref{gaintrans} and \eqref{noisyatt} that this channel family is realized by the attenuator-amplifier cascade in Eq.~\eqref{gencomp} with
\begin{align} \label{npsetag}
\eta(\kappa) &= \frac{\kappa}{N_B +1}, \\
G\pars{\kappa} &= N_B +1.
\end{align}
From Theorem~\ref{th:qfimub}, we immediately have that the QFI
\begin{align} \label{npsqfiub}
\cl{K}_{\kappa} \leq \widetilde{\cl{K}}_{\kappa} = \frac{N}{\kappa\pars{N_B + 1 -\kappa}}
\end{align}
which coincides with the bound derived in ref.~\cite{NG20} (cf. Eq.~(19) therein). The NPS property is faithfully captured by the lack of a modal  contribution to the QFI upper bound. 

For a fixed $N$, the upper bound \eqref{npsqfiub} can be closely approached using multimode independent and identically distributed (iid) two-mode squeezed vacuum (TMSV) probes of signal energy $N_S = N/M$ in the limit $M \gg 1$. For $\kappa \ll 1$, this is seen already from an approximate expression for the QFI of TMSV probes derived in \cite{SLHG-R+17}. For arbitrary values of $\kappa$, the  QFI of iid TMSV probes has been calculated recently \cite{JDC22,GRG+23} to be
\begin{align} \label{npsqfitmsv}
\cl{K}_{\kappa}[\mr{TMSV}]= N\,\frac{N_B + 1 + N_S\pars{N_B + 1 -\kappa}} {{\kappa \pars{N_B+1-\kappa} \bracs{N_B + 1 +N_S\pars{2N_B + 1 -\kappa}}},
}\end{align}
which approaches Eq.~\eqref{npsqfiub} in the limit of $N_S \rightarrow 0, M \rightarrow \infty$ with constant $N$. Various receiver designs realizing this performance were also analyzed in detail in Ref.~\cite{GRG+23}. We also mention that the optimal QFI attainable by {classical} probes is upper-bounded by the coherent-state value \cite{NG20}:
\begin{align} \label{npsqfics}
\cl{K}_{\kappa}[\mr{classical}] \leq \frac{N}{\kappa\pars{2N_B +1}},
\end{align}
 and is realized by homodyne detection in the $N_B \gg 1$ regime \cite{GRG+23}.

\subsection{Sensing with a Passive Signature} \label{sec:ps}

While the NPS assumption can be useful in the $\kappa \ll 1$ regime relevant to standoff sensing, it is problematic for situations of even moderately large values of $\kappa$ such as those that arise in laboratory settings, e.g., in quantum reading \cite{Pir11} or imaging \cite{GMT+20}. In such problems, varying $\kappa$ can result in substantial variations in the energy appearing at the output even when a vacuum probe is used -- this effect has been variously called  the `metrological power of the vacuum' \cite{PBG+18} or the `shadow effect'  \cite{JDC22}. We now consider transmittance estimation in this `\emph{passive signature (PS)}' regime.

To apply Theorem~\ref{th:qfimub}, we again let $K=1$ and $\theta_1 = \kappa$. From Eq.~\eqref{noisyatt}, we obtain the parameters
\begin{align}\label{psetag}
\eta\pars{\kappa} &= \frac{\kappa}{\pars{1-\kappa}N_B + 1}, \\
G\pars{\kappa} &= \pars{1-\kappa}N_B + 1,
\end{align}
for the attenuator-amplifier cascade generating the channel family $\left\{\cl{L}_{\kappa,N_B}\right\}$. For any probe with average energy $N$ in its $M$ signal modes, Theorem~\ref{th:qfimub}  gives the  upper bound
\begin{align} \label{psqfiub}
\cl{K}_{\kappa} \leq \widetilde{\cl{K}}_{\kappa} = N\,\frac{\pars{1+\kappa^2}N_B + 1}{\kappa\pars{1-\kappa}\bracs{\pars{1-\kappa}N_B + 1}^2} +
M\, \frac{N_B}{\pars{1-\kappa}\bracs{\pars{1-\kappa}N_B + 1}},
\end{align}
in terms of the photon and modal contributions respectively. 

The performance of single-mode and entangled two-mode Gaussian states for this problem has been studied in the works \cite{MI10,MI11,JDC22}. An exact expression for the QFI for an $M$-mode TMSV probe with per-mode signal energy $N_S = N/M$ is available from ref.~\cite{JDC22} (See Eq.~(36) therein - note that their expression needs to be reparametrized since our $\kappa$ corresponds to $\eta^2$ in \cite{JDC22}):
\begin{equation}\label{pstmsvqfi}
\begin{aligned} 
\cl{K}_{\kappa}[\mr{TMSV}] =  N\, \frac{\pars{1-\kappa}N_S + 2\kappa N_B +1}{\kappa\pars{1-\kappa}\bracs{\pars{1-\kappa}\pars{N_S+N_B+2N_SN_B} + 1}}\\
+
M\, \frac{N_B}{\pars{1-\kappa}\bracs{\pars{1-\kappa}\pars{N_S+N_B+2N_SN_B} + 1}}.
\end{aligned}
\end{equation}
Although we have separated  $\cl{K}_{\kappa}[\mr{TMSV}]$ into two terms to aid comparison with Eq.~\eqref{psqfiub}, it should be noted that, unlike our bound which splits into per-mode and per-photon contributions, both terms in Eq.~\eqref{pstmsvqfi} depend on $M$ and $N$.
The QFI of an $M$-mode coherent state of total energy $N$ is (see Eq.~(19) of \cite{JDC22})
\begin{equation}\label{pscsqfi}
\cl{K}_{\kappa}[\mr{coh}] =  N\, \frac{1}{\kappa\bracs{2\pars{1-\kappa}N_B + 1}}\\
+
M\, \frac{N_B}{\pars{1-\kappa}\bracs{\pars{1-\kappa}N_B + 1}},
\end{equation}
which can be shown using a standard convexity argument (See, e.g., \cite{NG20,NTG22})
to be the optimum QFI among all classical probes obeying the same constraints.

\begin{figure*}[t] 
    \begin{subfigure}[b]{0.345\textwidth}\label{figNBfig}
    \centering
    {\includegraphics[trim=8mm 62mm 23mm 67mm, clip=true, scale=0.34]{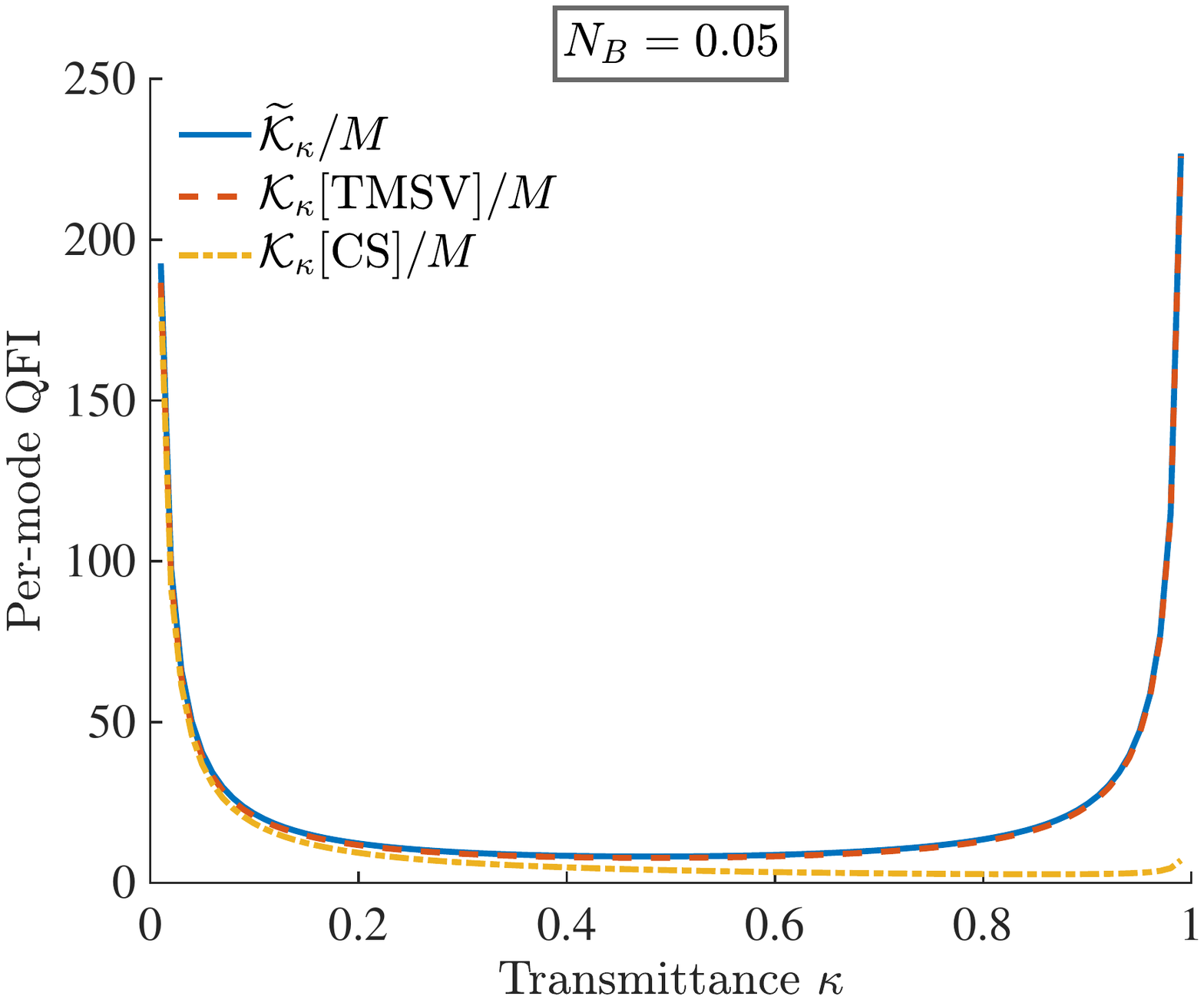}} 
    \end{subfigure}
     \begin{subfigure}[b]{0.322\textwidth}
    \centering
    {\includegraphics[trim=20mm 62mm 21mm 67mm, clip=true, scale=0.34]{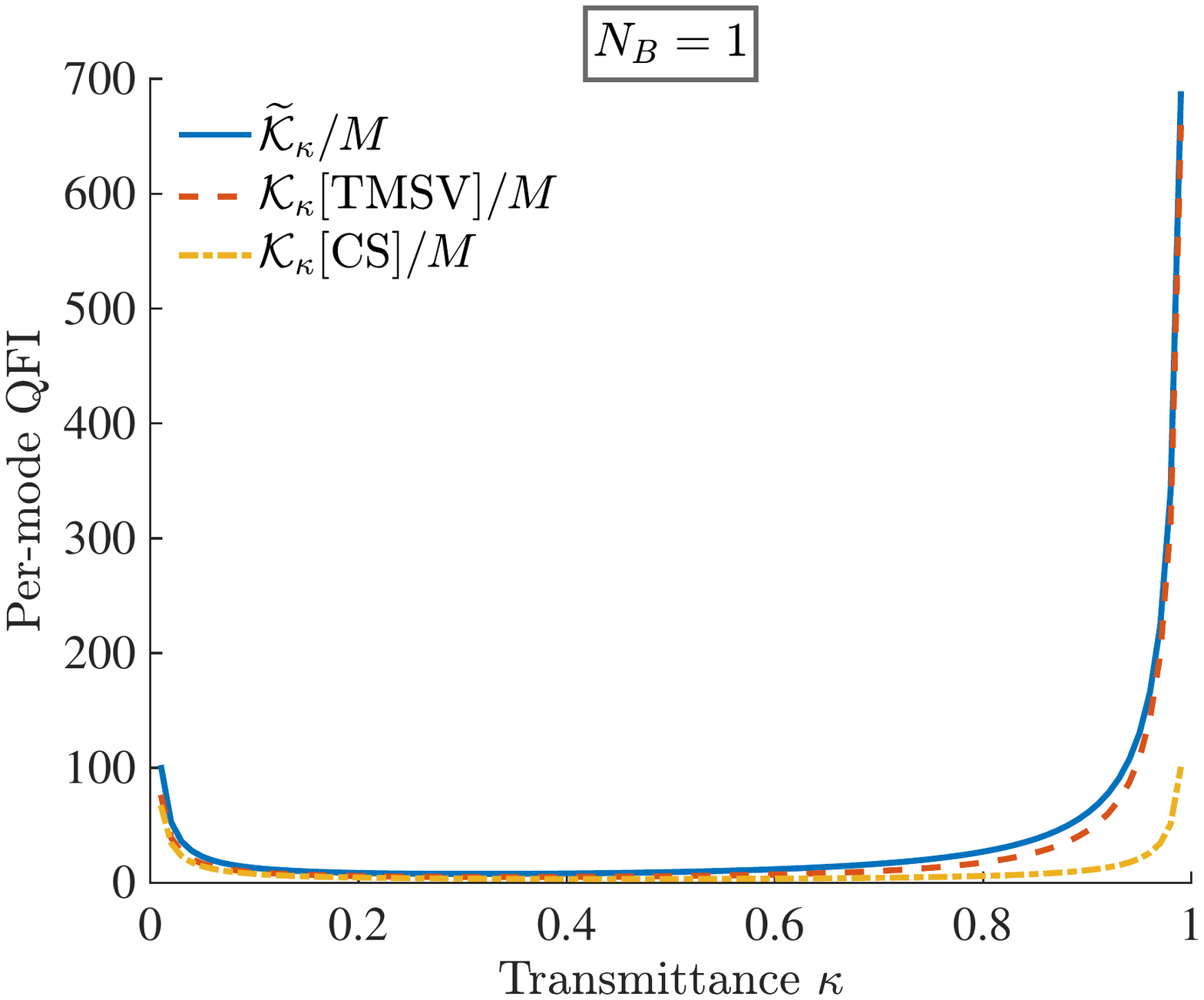}} 
    \end{subfigure}
     \begin{subfigure}[b]{0.322\textwidth}
    \centering
    {\includegraphics[trim=19mm 62mm 23mm 67mm, clip=true, scale=0.34]{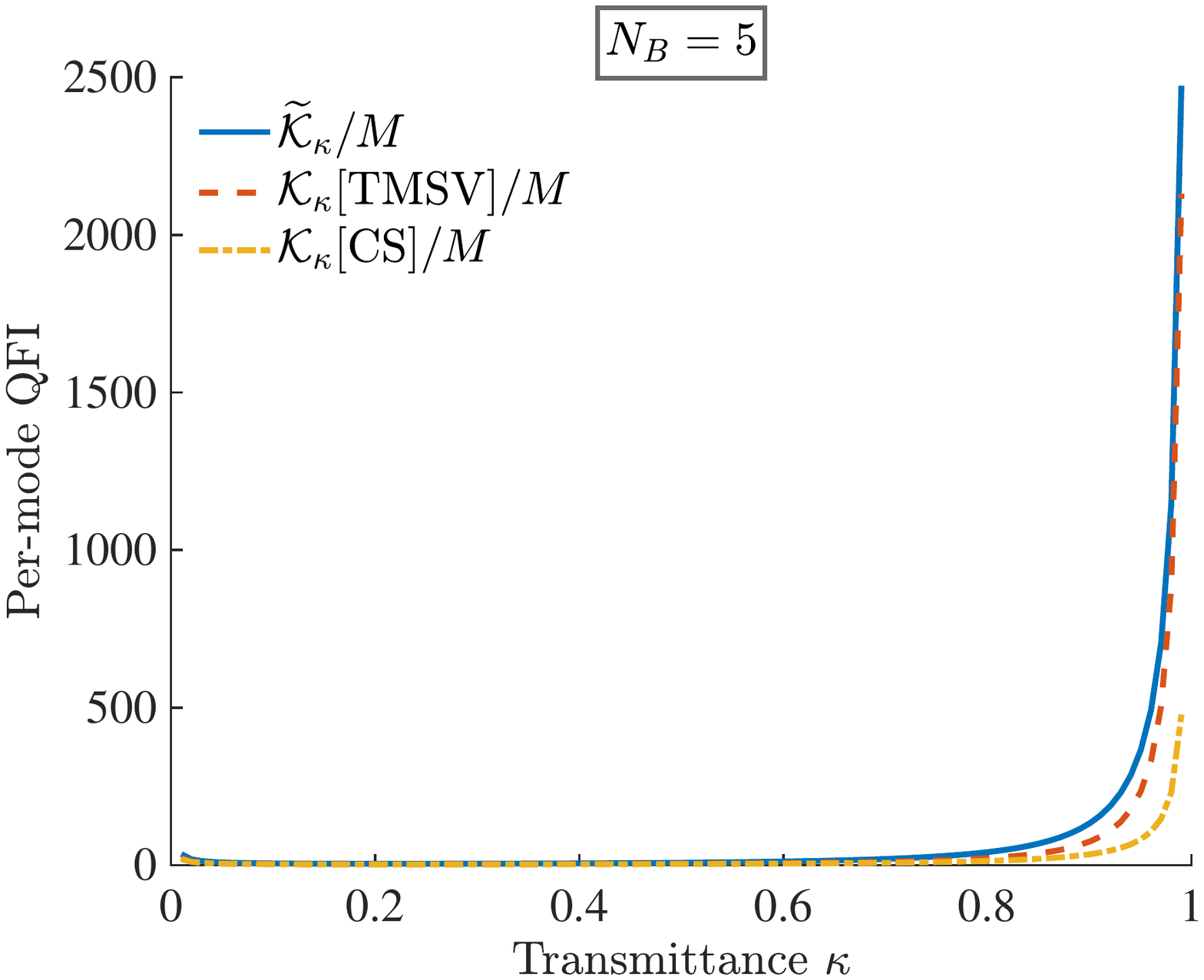}} 
    \end{subfigure}
    \caption{Dependence of the QFI for $\kappa$ on the noise brightness $N_B$: Comparison of the per-mode QFI of iid TMSV probes (Eq.~\eqref{pstmsvqfi} (red dashed)), coherent-state probes (Eq.~\eqref{pscsqfi} (yellow dashed-dotted)) and the universal upper bound $\widetilde{\mathcal{K}}_{\kappa}/M$ (Eq.~\eqref{psqfiub} (blue solid)) for (Left) $N_B = 0.05$, (Center) $N_B = 1$, and (Right) $N_B=5$. $M=5$ and $N=10$ in all the plots.}\label{fig:NBfig}
\end{figure*}

While the above expressions are rather complicated, it is useful to examine some limiting cases. Setting $N_B = 0$ recovers the known performances of TMSV and coherent-state probes for sensing  pure-loss channels \cite{Nai18loss}. For $N_B>0,$ setting $N=N_S=0$ recovers the vacuum-probe performance in all three expressions.  

In Fig.~\ref{fig:NBfig}, we compare these QFI expressions for different values of $N
_B$ at fixed values of $M=5$ and $N=10$. For $N_B \sim 0$, the TMSV probe performs much better than classical probes in the $\kappa \sim 1$ regime, similar to the behavior of noiseless loss sensing \cite{Nai18loss}. As $N_B$ increases, the coherent-state QFI begins to rise in the $\kappa \sim1$ regime due to the passive signature effect acting through the modal contribution in Eq.~\eqref{pscsqfi}. Importantly, however, we note that the TMSV performance is throughout close to the upper bound \eqref{psqfiub}, making it near-optimal among all quantum probes.

\begin{figure*}[t] 
    \begin{subfigure}[b]{0.345\textwidth}
    \centering
    {\includegraphics[trim=8mm 62mm 23mm 67mm, clip=true, scale=0.34]{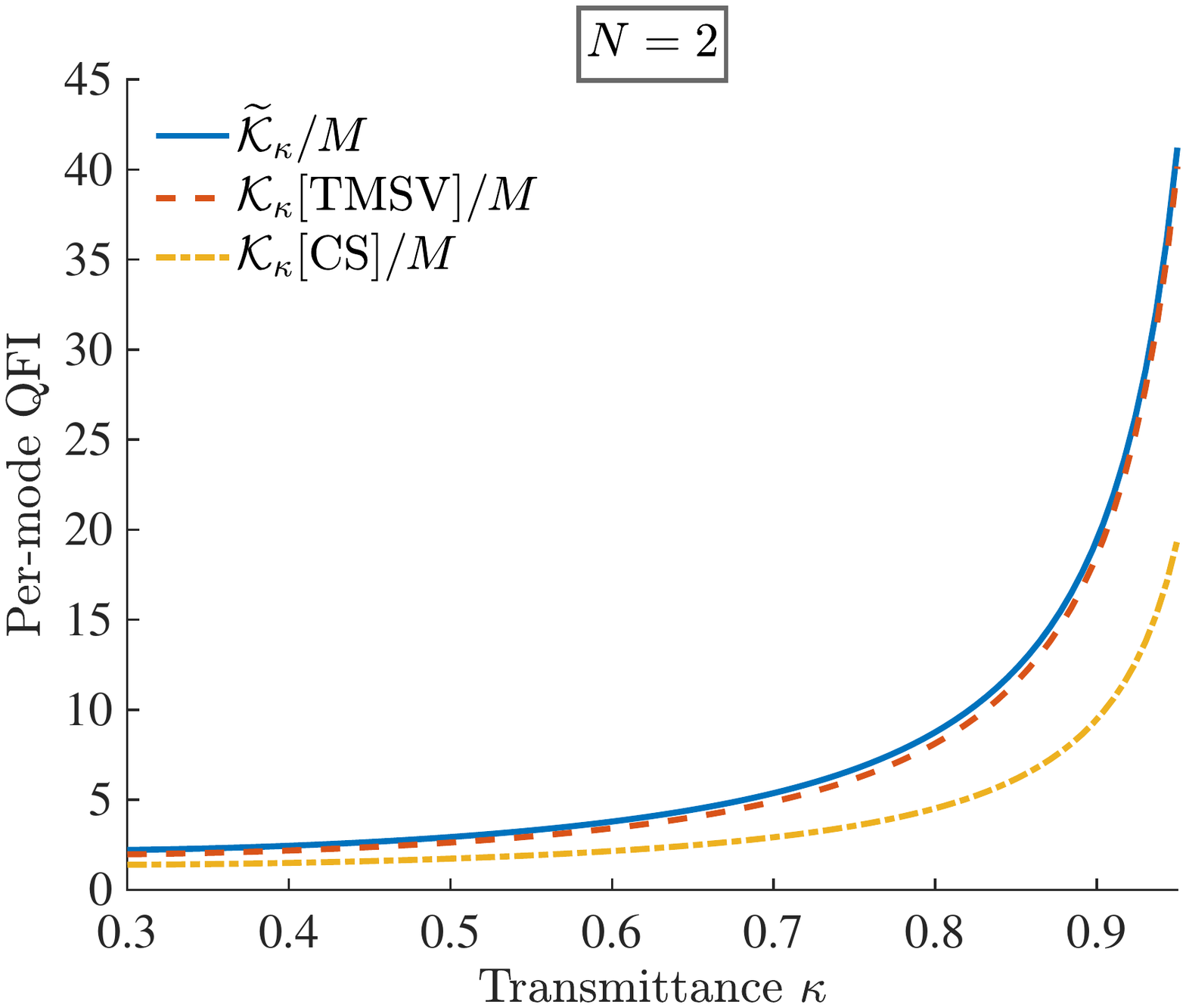}} 
    \end{subfigure}
     \begin{subfigure}[b]{0.322\textwidth}
    \centering
    {\includegraphics[trim=23mm 62mm 21mm 67mm, clip=true, scale=0.34]{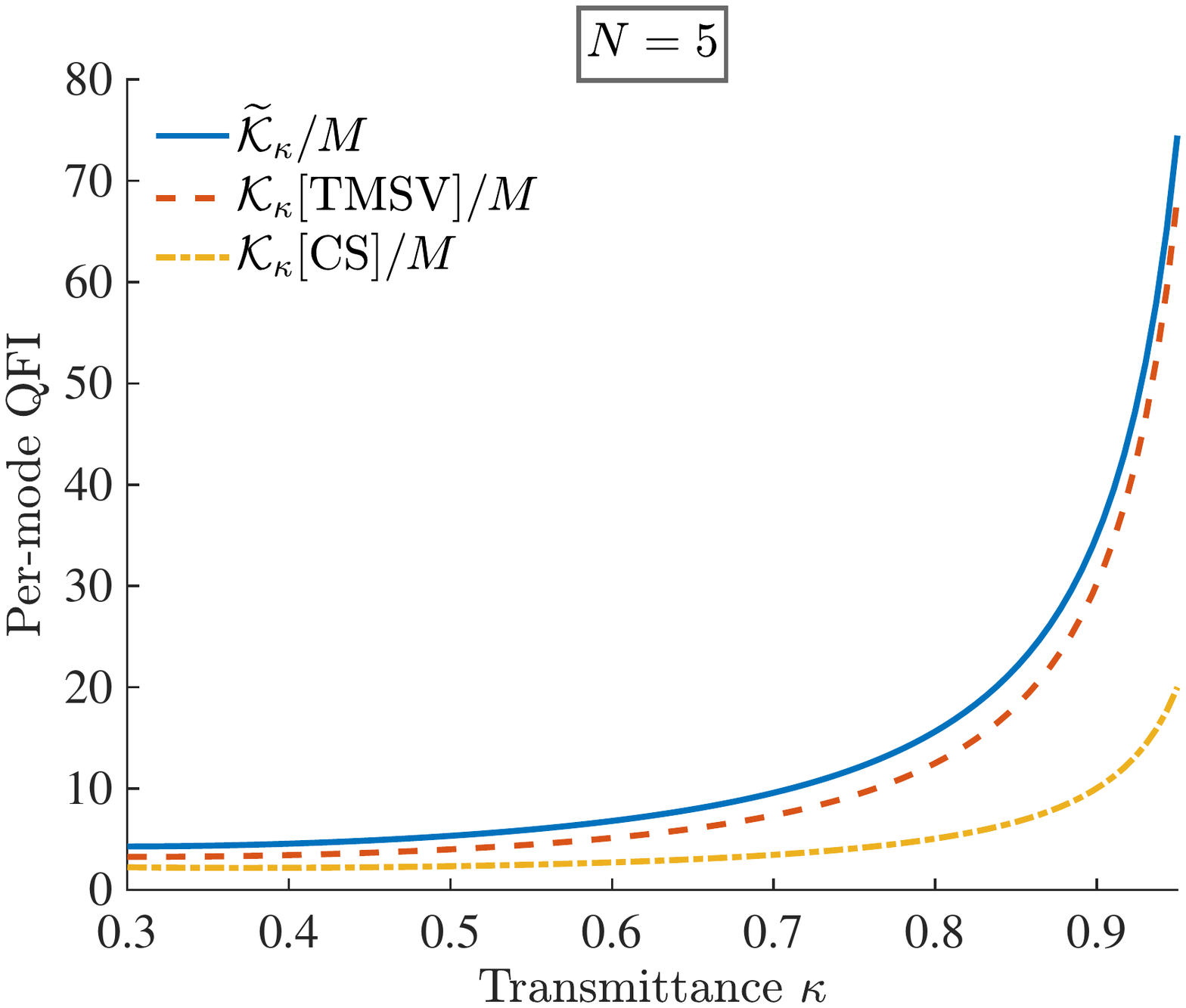}} 
    \end{subfigure}
     \begin{subfigure}[b]{0.322\textwidth}
    \centering
    {\includegraphics[trim=20mm 62mm 23mm 67mm, clip=true, scale=0.34]{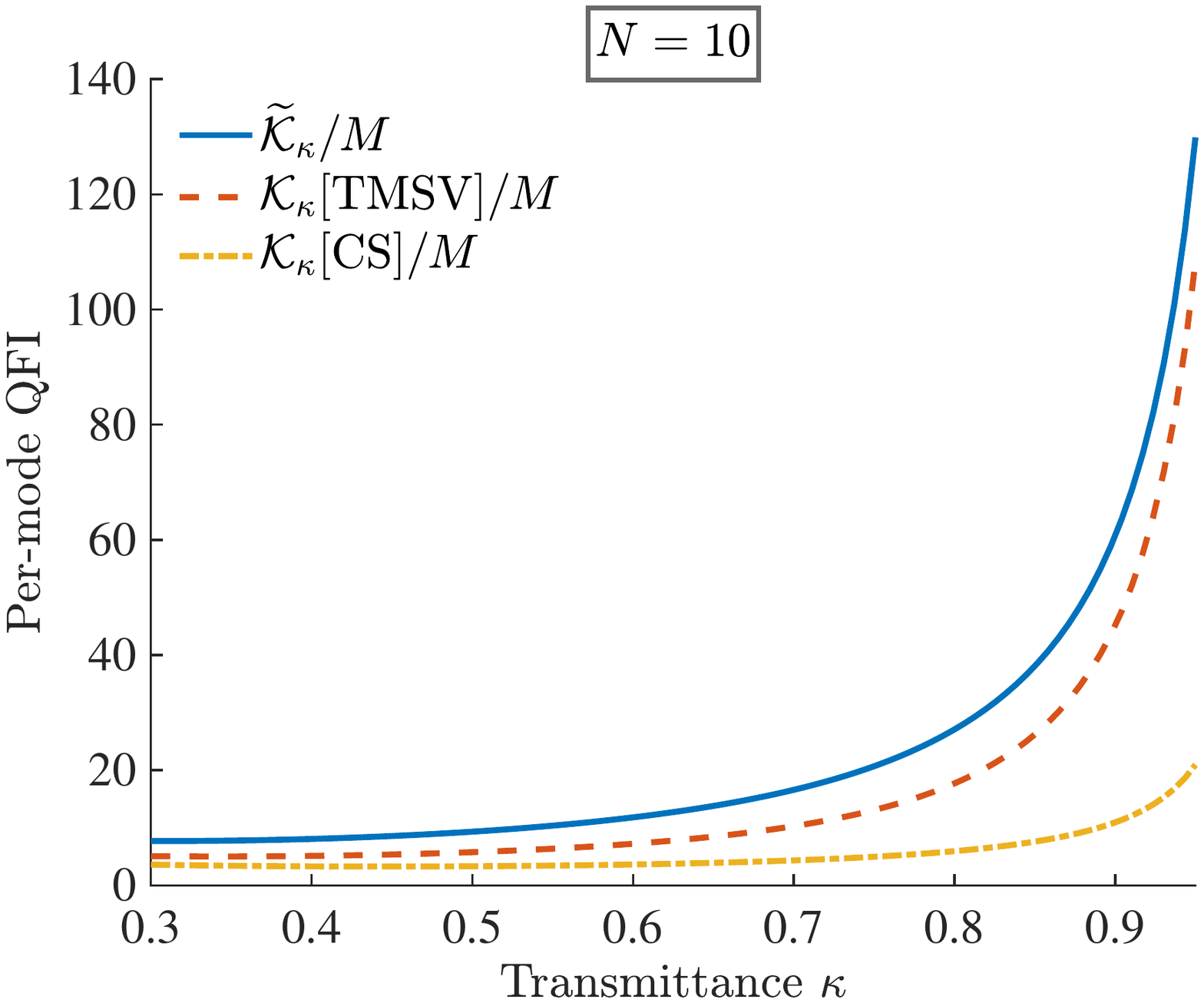}} 
    \end{subfigure}
    \caption{Dependence of the QFI for $\kappa$ on the signal energy $N$: Comparison of the per-mode QFI of iid TMSV probes (Eq.~\eqref{pstmsvqfi} (red dashed)), coherent-state probes (Eq.~\eqref{pscsqfi} (yellow dashed-dotted)) and the universal upper bound $\widetilde{\mathcal{K}}_{\kappa}/M$ (Eq.~\eqref{psqfiub} (blue solid)) for (Left) $N = 2$, (Center) $N = 5$, and (Right) $N=10$. $M=5$ and $N_B=1$ in all the plots.}\label{fig:Nfig}
\end{figure*}

In Fig.~\ref{fig:Nfig}, we compare our bound with the TMSV and coherent-state performance for different values of $N$ at fixed values of $M=5$ and $N_B=1$. As $N$ is increased, the absolute magnitude of the QFI increases as well as the relative advantage of TMSV over coherent states, especially in the $\kappa \sim 1$ regime.  While the TMSV performance is still close to the upper bound \eqref{psqfiub}, it becomes even more so as the signal brightness $N_S=N/M$ decreases.

\begin{figure*}[h] 
    \begin{subfigure}[b]{0.345\textwidth}
    \centering
    {\includegraphics[trim=8mm 62mm 23mm 67mm, clip=true, scale=0.34]{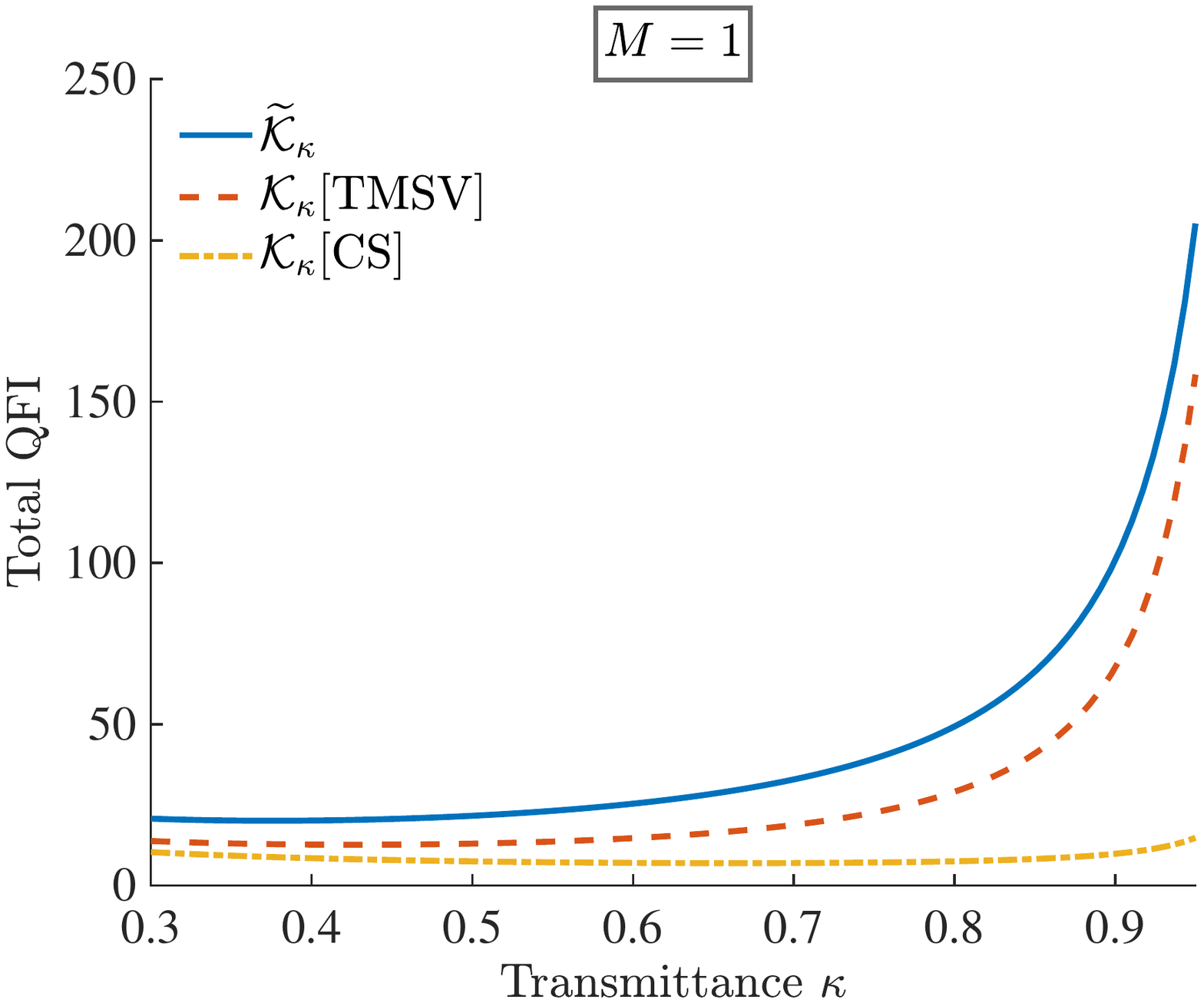}} 
    \end{subfigure}
     \begin{subfigure}[b]{0.322\textwidth}
    \centering
    {\includegraphics[trim=22mm 62mm 21mm 67mm, clip=true, scale=0.34]{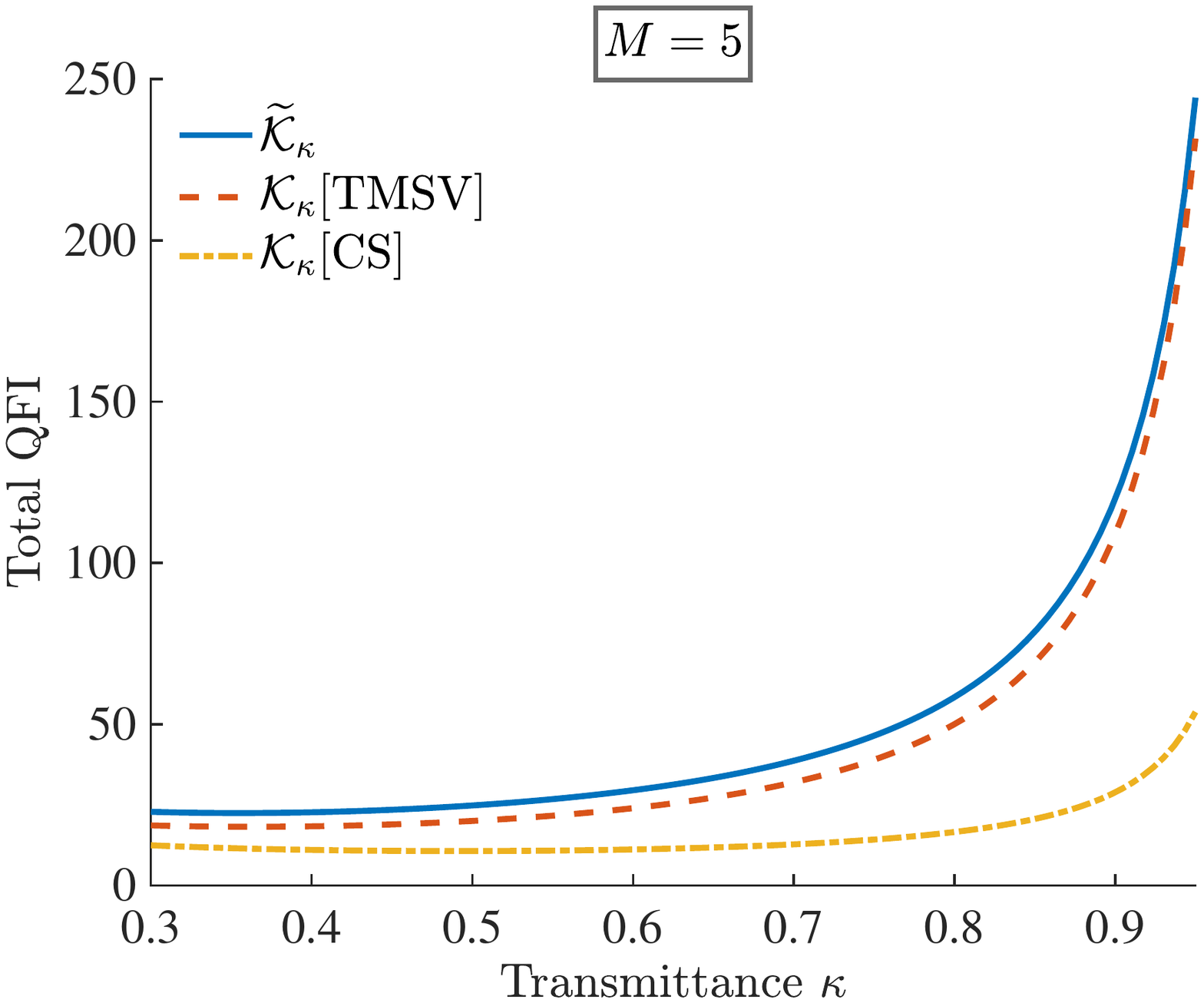}} 
    \end{subfigure}
     \begin{subfigure}[b]{0.322\textwidth}
    \centering
    {\includegraphics[trim=20mm 62mm 23mm 67mm, clip=true, scale=0.34]{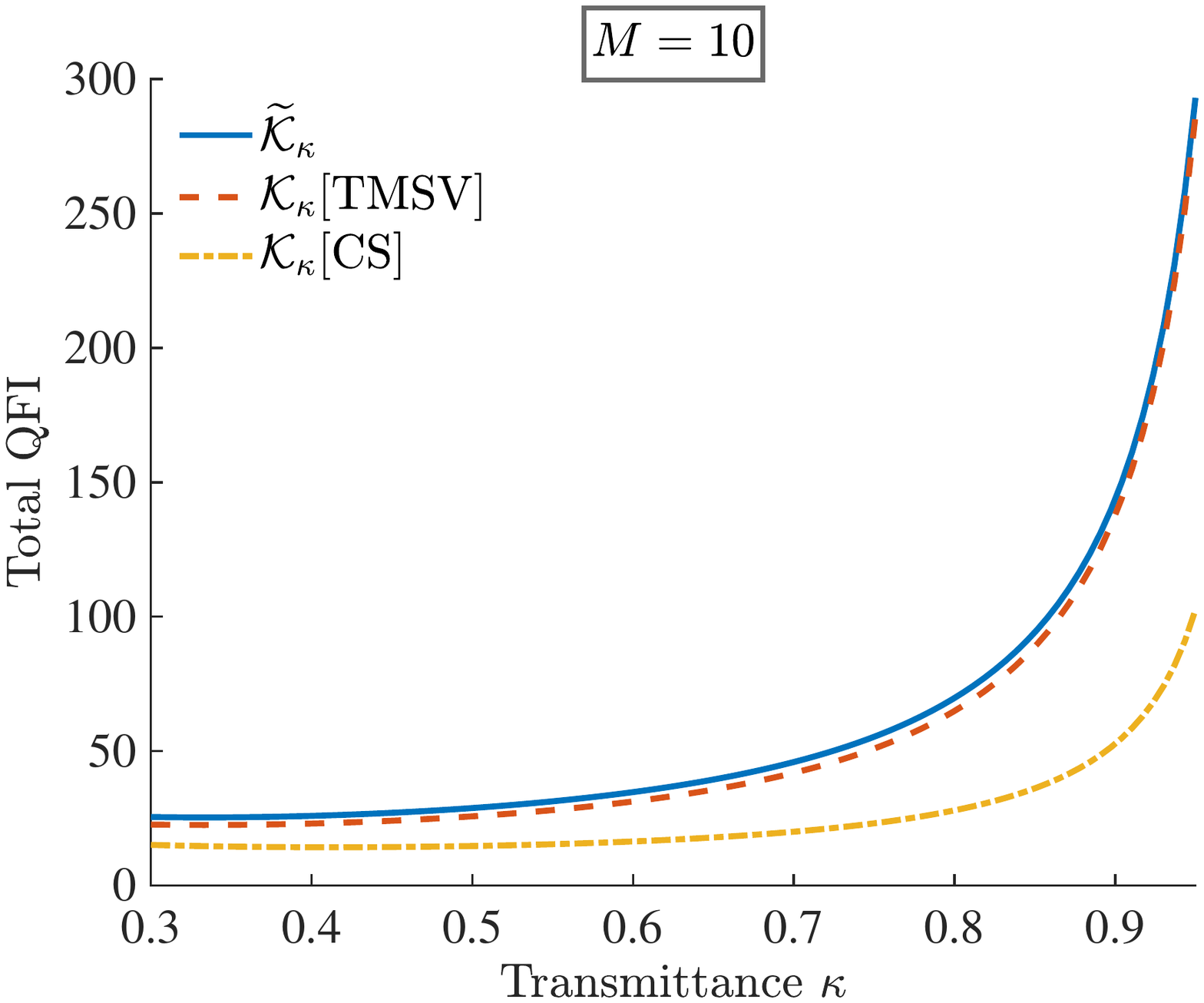}} 
    \end{subfigure}
    \caption{Dependence of the QFI for $\kappa$ on the number of modes $M$: Comparison of the per-mode QFI of iid TMSV probes (Eq.~\eqref{pstmsvqfi} (red dashed)), coherent-state probes (Eq.~\eqref{pscsqfi} (yellow dashed-dotted)) and the universal upper bound $\widetilde{\mathcal{K}}_{\kappa}/M$ (Eq.~\eqref{psqfiub} (blue solid)) for (Left) $M = 1$, (Center) $M = 5$, and (Right) $M=10$. $N=5$ and $N_B=0.5$ in all the plots.}\label{fig:Mfig}
\end{figure*}

In Fig.~\ref{fig:Mfig}, we compare our bound with the TMSV and coherent-state performance for different values of $M$ at fixed values of $N=5$ and $N_B=0.5$. Already at $M=5$, we find that the TMSV probe is near-quantum-optimal for any $\kappa$ and has a large advantage over coherent states for $\kappa \sim 1$. The TMSV performance improves with increasing $M$ (thus decreasing the signal brightness $N_S$), mimicking the behavior under the NPS assumption.

From a practical standpoint, all the plots indicate that realizing a substantial quantum advantage  over classical sensing with TMSV probes requires one to be sensing in the regime of $\kappa \gtrsim 0.7$.

Finally, we mention that for $M=1$, our upper bound \eqref{psqfiub} coincides with one derived for  single-mode probes by Wang et al. \cite{WDA20} using a purification of the thermal loss channel (See  Eq.~(A7) therein under the reparametrization $\kappa = e^{-2\gamma t}$, where $\gamma$ is the damping rate and $t$ is the interaction time in their master equation formulation). The relationship of our bound to purification-based approaches is an interesting question for future work (Cf. Remark~\ref{rem1} above).

\section{Application: Sensing noise level of additive-noise channels} \label{sec:anlsensing}

The additive-noise (or `classical-noise') channel  \cite{Hol19qsci} is the continuous-variable quantum analog of the additive white Gaussian noise channel from classical communication theory. In the Heisenberg picture, a mode with annihilation operator $\hat{a}_\tsf{in}$ is transformed to the output annihilation operator
\begin{align} \label{hp}
\hat{a}_\tsf{out} = \hat{a}_\tsf{in} + n,
\end{align}
where $n=n_r + in_i$ is a complex random amplitude whose real and imaginary parts are independent and identically distributed zero-mean Gaussian random variables of variance $\gamma/2$ each, where $\gamma$ is the noise level of the channel. The resulting characteristic function transformation is
\begin{align}
\chi_\tsf{out}\pars{\xi} = \chi_\tsf{in}\pars{\xi}e^{-\gamma \abs{\xi}^2}.
\end{align} 
We denote the resulting quantum channel by $\cl{N}_{\gamma}$. The isotropic nature of the added noise makes $\cl{N}_{\gamma}$ phase-covariant. It arises operationally in continuous-variable quantum teleportation \cite{FSB+98} owing to the shared entangled state having only a finite degree of squeezing \cite{BK98}.  The advantage of using quantum (rather than coherent-state) probes to distinguish between an ideal teleportation channel $\cl{N}_0$ (which is just the identity channel) and an imperfect implementation $\cl{N}_{\gamma}$  has been studied by many authors -- see, e.g., \cite{CW04,SSW22} and references therein. Recently, Sharma et al. obtained  ancilla-entangled probes with $M=1$ that minimize the fidelity between the outputs of two such channels under an energy constraint \cite{SSW22}. These works presume a prior estimate of the actual noise level $\gamma$ of the imperfect implementation. Here, we apply Theorem~\ref{th:qfimub} to find a limit on the precision in estimating this noise level using ancilla-entangled $M$-signal-mode probes under a signal energy constraint of $N$ photons.

\begin{figure}[t] 
    \begin{subfigure}[b]{0.46\textwidth}
    \centering
      \includegraphics[trim=5mm 50mm 10mm 55mm, clip=true, scale=0.44]{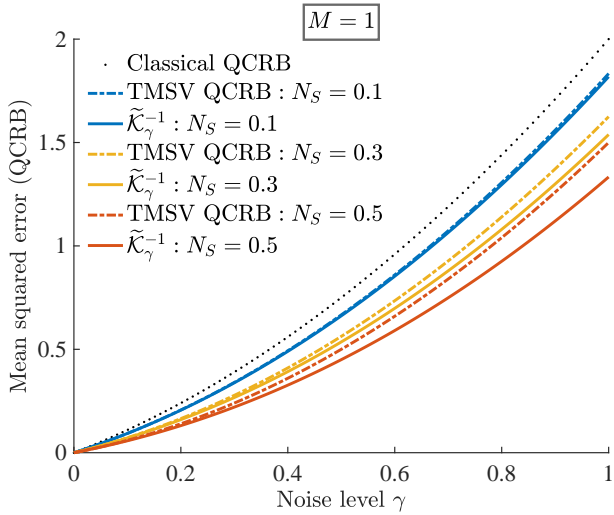}
  \subcaption{QCRBs for classical probes (dotted), for TMSV probes (dashed-dotted), and our universal lower bound  on the mean squared error (solid) for sensing the noise level of an additive-noise channel for signal brightness $N_S = 0.1$ (blue), $N_S = 0.3$ (yellow), and $N_S = 0.5$ (red). $M=1$ for all the plots.}
  \label{fig:nlndep}
    \end{subfigure}
    \hfill
     \begin{subfigure}[b]{0.46\textwidth}
    \centering
   \includegraphics[trim=5mm 50mm 10mm 55mm, clip=true, scale=0.41]{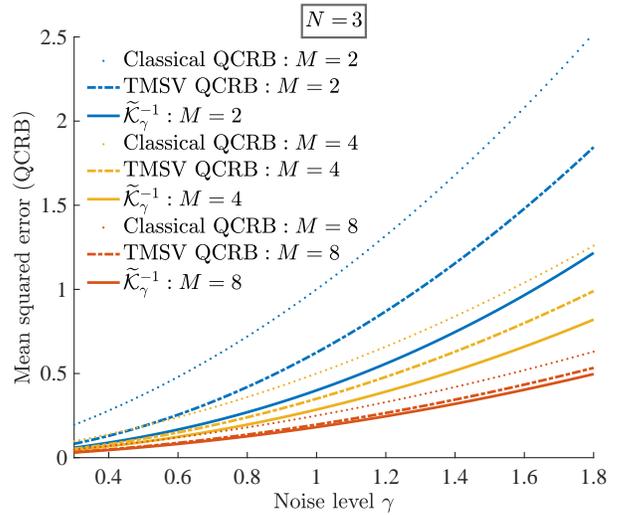}
  \subcaption{QCRBs for classical probes (dotted), for TMSV probes (dashed-dotted) and our universal lower bound on the mean squared error (solid) for sensing the noise level of an additive-noise channel for signal brightness $M = 2$ (blue), $M = 4$ (yellow), and $M = 8$ (red). The total signal energy $N=3$ for all the plots.}
  \label{fig:nlmdep}
  \end{subfigure}
  \caption{Sensing noise level of an additive-noise channel: QCRBs for classical and TMSV probes and the universal mean squared error bound from Eq.~\eqref{nlqfiub}.}
\end{figure}

As before, it is readily shown that $\cl{N}_{\gamma}$ admits  decomposition into an attenuator-amplifier cascade with the parameters
\begin{align}
\eta\pars{\gamma} &= \pars{\gamma +1}^{-1}, \\
G\pars{\gamma} &= \gamma +1.
\end{align}
Setting $K=1$ and $\theta_1 = \gamma$ in Theorem~\ref{th:qfimub} gives the upper bound
\begin{align} \label{nlqfiub}
\cl{K}_{\gamma} \leq  \widetilde{\cl{K}}_{\gamma} = \frac{2N}{\gamma(\gamma+1)^2} + \frac{M}{\gamma(\gamma+1)}, 
\end{align}
which exhibits both a photon and a modal contribution (Cf.~\ref{rem3}).

The modal contribution above is in fact the best QFI that \emph{any} classical probe can achieve. To see this, note that $\cl{N}_{\gamma}$ commutes with an arbitrary displacement, so any displacement at the input can be undone at the output. Therefore, the performance of any coherent-state probe (and hence, any classical probe) is the same as that of a vacuum probe. Since $\cl{N}_{\gamma}\pars{\ket{0}\bra{0}} = \rho_{\tsf{th}}\pars{\gamma}$, the problem of estimating $\gamma$ using an $M$-signal-mode classical probe reduces to that of estimating the average energy of a thermal state from $M$ copies, which is well-known to obey the QFI
\begin{align} \label{nlcsqfi}
\cl{K}_{\gamma}[\mathrm{CS}] = \frac{M}{\gamma(\gamma+1)}
\end{align}
that is achieved by photon counting \cite{Hel76}.
Thus, we must turn to quantum probes to realize any $N$-dependent precision enhancement. Consider as a probe $M$ iid copies of a TMSV state with per-mode brightness $N_S = N/M$. Since the output states are Gaussian, we leverage existing techniques (e.g, see \cite{BBP15,SLF15}) to evaluate the QFI:
\begin{align} \label{tmsvqfi}
\cl{K}_{\gamma}\bracs{\mathrm{TMSV}} = \frac{2N+M}{\gamma\bracs{(2N_S+1)\gamma + 1}}.
\end{align}
We see that the above expression agrees with Eq.~\eqref{nlcsqfi} and the upper bound \eqref{nlqfiub} in the vacuum limit $N_S \rightarrow 0$ for any $M$. It can also be verified that $
\cl{K}_{\gamma}\bracs{\mathrm{TMSV}}  \leq \widetilde{\cl{K}}_{\gamma}$ for all $N$ and $M$.
The resulting QCRBs for the TMSV probe, the universal lower bound on the MSE from Eq.~\eqref{nlqfiub}, and the classical performance from Eq.~\eqref{nlcsqfi} are plotted in Fig.~\ref{fig:nlndep} and three values of $N$. We note that the TMSV performance is close to the universal bound especially in the regime of small $\gamma$ and that they both yield better precision with increasing energy. The agreement between the TMSV performance and the bound increases as the brightness $N_S$ is decreased. This effect is also seen in Fig.~\ref{fig:nlmdep} in which $M$ is increased while keeping $N$ fixed.

However,  the TMSV performance in Eq.~\eqref{tmsvqfi} begins to diverge from our bound in Eq.~\eqref{nlqfiub} in the high-brightness regime. Indeed, if $M$ is kept fixed and $N$ (and therefore, $N_S$) is increased, the TMSV performance saturates at $M/\gamma^2$ as $N \rightarrow \infty$. On the other hand, our bound Eq.~\eqref{nlqfiub} increases without limit. Indeed, we believe that -- unlike transmittance sensing -- arbitrarily large precision cannot be had in this problem by increasing $N$ no matter what probe is used, similar to the findings of Refs.~\cite{PL17,TW16arxiv} for sensing the noise brightness of a thermal loss channel. We leave further study of these issues for future work.

\section{Discussion} \label{sec:discussion}

In this paper, we have made two general contributions to the study of entanglement-assisted quantum sensing of optical channels. First, we have shown that  the search for optimal probes for any  sensing problem involving multiparameters of phase-covariant channels under energy and mode constraints can be limited to the class of number-diagonal signal (NDS) probes. Although this simplification has been explored previously for particular problems involving Gaussian channels \cite{KD-D10,Nai11,CDB+14,NY11,Nai18,Nai18loss,SWA+18,NTG22}, our Theorem \ref{th:ndsoptimality} enables it to be systematically exploited for quantum sensing. Beyond the Gaussian-channel framework, it can also be applied to sensing  channels such as phase-diffusion channels \cite{VDG+14,SBD17} and for studying quantum effects in multi-photon sensing schemes, e.g., sensing with two-photon absorption \cite{SMFS21}. 

Secondly, we have exploited the decomposition theorem for phase-covariant Gaussian channels to derive an easily computed upper bound on the quantum Fisher information matrix (QFIM) for sensing any such family of channels. As test cases, our bound readily yielded performance limits for sensing the transmittance of thermal loss channels and the noise variance of additive-noise channels. Similar to the case of standoff transmittance sensing using the no-passive-signature assumption \cite{SLHG-R+17,NG20,JDC22,GRG+23}, it was shown that iid two-mode squeezed vacuum (TMSV) probes become near-optimal in the limit of low per-mode brightness. For transmittance sensing, the optimality of low-brightness TMSV probes makes them attractive for probing delicate biological samples \cite{TB16}, while our results indicate that the quantum advantage is maximized in the regime of $\kappa \sim 1$, i.e., for nearly transparent samples.

Investigation of measurements that can harness the quantum advantages of low-brightness TMSV probes for these problems is clearly of interest. More generally, Theorem \ref{th:qfimub} can be applied to a multitude of sensing problems, e.g., the sensing the noise brightness of a thermal loss channel \cite{PL17,TW16arxiv,WDA20,SZ23}, joint sensing of the transmittance and noise brightness, and the sensing of parameters of linear amplifiers beyond the quantum-limited ones studied in ref.~\cite{NTG22}.

\section{Acknowledgements}
This work is supported by the Singapore Ministry of Education Tier 2 Grant No. T2EP50221-0014. Any opinions, findings and conclusions or recommendations expressed in this material are those of the author(s) and do not reflect the views of National Research Foundation or the Ministry of Education, Singapore.

%
%\newpage
\appendix
\section{Fidelity between  Output States of Amplifier Channels: Derivation of Eq.~\eqref{conditionalfid}} \label{app:appendix}

In order to get Eq.~\eqref{conditionalfid}, we first show the following general result. 
Consider an $M$-mode signal ($S$) system, an arbitrary ancilla ($A$) system, and two quantum-limited product amplifier channels $\cl{A}_{G}^{\otimes M}$ and $\cl{A}_{G'}^{\otimes M}$ acting on $S$. Suppose that an $NDS$ probe $\ket{\psi}_{AS} = \sum_{\mb{n} \geq \mb{0}} \sqrt{p_{\mb{n}}} \keta{\chi_{\mb{n}}}\kets{\mb{n}}$ is input to each of these channels, resulting in the output states $\pars{\mathrm{id}_A \otimes \cl{A}_{G}^{\otimes M}} \Psi$ and $\pars{\mathrm{id}_A \otimes \cl{A}_{G'}^{\otimes M}} \Psi$ respectively on $AS$, where $\Psi := \ket{\psi}\bra{\psi}_{AS}$. The fidelity between these output states was computed in Ref.~\cite{NTG22}.  The following theorem generalizes this result by computing the  fidelity between the output states of two quantum-limited amplifier channels on $S$  when their inputs are \emph{distinct} NDS states entangled in the same ancilla basis $\left\{\keta{\chi_{\mb{n}}}\right\}$.

\begin{thm} \label{thm:ampoutputfidelity}
With $\mb{n} = \pars{n_1,\ldots, n_M}$ indexing the $M$-mode Fock states of $S$, consider the NDS states
\begin{equation} \label{states}
\begin{aligned}
\ketas{\psi} &= \sum_{\mb{n}} \sqrt{r_\mb{n}} \keta{\chi_\mb{n}} \kets{\mb{n}}, \\
\ketas{\psi'} &= \sum_{\mb{n}} \sqrt{s_\mb{n}} \keta{\chi_\mb{n}} \kets{\mb{n}},
\end{aligned}
\end{equation}
where $\left\{r_\mb{n}\right\}$ and  $\left\{s_\mb{n}\right\}$ are arbitrary multimode photon probability distributions and $\left\{\keta{\chi_\mb{n}} \right\}$ is a given orthonormal set of states in $A$.  Suppose that
\begin{equation}
\begin{aligned}
\rho &= \pars{\mr{id}_A \otimes \cl{A}_G^{\otimes M } }\pars{\ket{\psi}\bra{\psi}}, \\
\rho' &= \pars{\mr{id}_A \otimes \cl{A}_{G'}^{\otimes M } }\pars{\ket{\psi'}\bra{\psi'}}.
\end{aligned}
\end{equation}
The fidelity between these  output states is given by
\begin{align}
F\pars{\rho, \rho'} &= \sum_{\mb{n} \geq \mb{0}}  \sqrt{r_\mb{n}\, s_\mb{n}}\;\nu^{n+M}, 
\end{align}
where $n = \sum_{m=1}^M n_m$ 
and 
\begin{equation} \label{nudef}
\nu = \sech(\tau'-\tau) = \pars{\sqrt{GG'} - \sqrt{\pars{G-1}\pars{G'-1}}}^{-1}\in (0,1],
\end{equation}
in terms of the parametrization  $G = \cosh^2 \tau, G' = \cosh^2 \tau'$ \cite{NTG22}.
\begin{proof}
The single-mode quantum-limited amplifier map can be written in Kraus form \cite{ISS11} as 
\begin{align}
\cl{A}_G \pars{\sigma} &= \sum_{a=0}^\infty \hat{K}_a(\tau) \,\sigma \,\hat{K}_a^{\dag}(\tau),  \\
\hat{K}_a (\tau)&=  \bracs{\tanh \tau}^a \sum_{k=0}^\infty \sqrt{{k+a \choose a}} \bracs{\sech \tau}^{k+1} \ket{k+a}\bra{k}, \label{amplifierKrausops}
\end{align}
and $a \geq 0$ can be interpreted as the number of photons added to the $S$ mode during the two-mode squeezing interaction between the input mode and internal mode of the amplifier. We thus have, for $\mb{a} = \pars{a_1,\ldots, a_M} \geq \mb{0}$,
\begin{align}
\rho_{AS} &= \sum_{\mb{a} \geq \mb{0}} \varket{\Psi_\mb{a}}\varbra{\Psi_\mb{a}},\\
\rho'_{AS} &= \sum_{\mb{a} \geq \mb{0}} \varket{\Psi'_\mb{a}}\varbra{\Psi'_\mb{a}},
\end{align}
where 
\begin{align}
\varket{\Psi_\mb{a}} &= \hat{I}_A \otimes \pars{\bigotimes_{m=1}^M \hat{K}_{a_m}(\tau)} \ket{\Psi}, \\
\varket{\Psi'_\mb{a}} &= \hat{I}_A \otimes \pars{\bigotimes_{m=1}^M \hat{K}_{a_m}\pars{\tau'}} \ket{\Psi'}
\end{align}
are unnormalized kets. Using Eq.~\eqref{amplifierKrausops}, we find that
\begin{align}
\varket{\Psi_\mb{a}} &= \bracs{\tanh \tau}^{\tr \mb{a}} \sum_{\mb{n} \geq \mb{0}} \sqrt{r_\mb{n}}  \bracs{\sech \tau}^{\tr \mb{n} +M} \sqrt{\prod_{m=1}^M {n_m + a_m \choose a_m}} \keta{\chi_\mb{n}}\pars{\bigotimes_{m=1}^M \kets{n_m + a_m}}, \\
\varket{\Psi'_\mb{a}} &= \bracs{\tanh \tau'}^{\tr \mb{a}} \sum_{\mb{n} \geq \mb{0}} \sqrt{s_{\mb{n}}}  \bracs{\sech \tau'}^{\tr \mb{n} +M} \sqrt{\prod_{m=1}^M {n_m + a_m \choose a_m}} \keta{\chi_\mb{n}}\pars{\bigotimes_{m=1}^M \kets{n_m + a_m}}.
\end{align}
We thus have
\begin{align}
\varbraket{\Psi_\mb{a}}{\Psi'_\mb{a'}} = \delta_{\mb{a},\mb{a'}}\,\bracs{\tanh \tau \cdot \tanh \tau'}^{\tr \mb{a}} \sum_{\mb{n} \geq \mb{0}} \sqrt{r_\mb{n}\,s_\mb{n}}  \bracs{\sech \tau \cdot \sech \tau'}^{\tr \mb{n} +M} \bracs{\prod_{m=1}^M {n_m + a_m \choose a_m}},
\end{align}
where the orthonormality of $\left\{\keta{\chi_\mb{n}}\right\}$ has been used. The fact that $\Psi_\mb{a}$ and $\Psi'_\mb{a'}$ are orthogonal for $\mb{a} \neq \mb{a'}$ allows us to compute the fidelity
\begin{align}
&F(\rho,\rho') := \Tr \sqrt{\sqrt{\rho} \,\rho' \sqrt{\rho}} 
= \sum_{\mb{a} \geq \mb{0}} \abs{\varbraket{\Psi_\mb{a}}{\Psi'_\mb{a'}}} 
= \sum_{\mb{a} \geq \mb{0}} \varbraket{\Psi_\mb{a}}{\Psi'_\mb{a'}} \\
&= \sum_{\mb{a} \geq \mb{0}} \bracs{\tanh \tau \cdot \tanh \tau'}^{\tr \mb{a}} \sum_{\mb{n} \geq \mb{0}} \sqrt{r_\mb{n}\,s_\mb{n}}  \bracs{\sech \tau \cdot \sech \tau'}^{\tr \mb{n} +M} \bracs{\prod_{m=1}^M {n_m + a_m \choose a_m}} \\
&= \sum_{\mb{n} \geq \mb{0}} \bracs{\sech \tau \cdot \sech \tau'}^{\tr \mb{n} +M} \sqrt{r_\mb{n}\,s_\mb{n}} \sum_{\mb{a} \geq \mb{0}} \bracs{\tanh \tau \cdot \tanh \tau'}^{\tr \mb{a}} \bracs{\prod_{m=1}^M {n_m + a_m \choose a_m}}\\
&= \sum_{\mb{n} \geq \mb{0}} \bracs{\sech \tau \cdot \sech \tau'}^{\tr \mb{n} +M} \sqrt{r_\mb{n}\,s_\mb{n}} \sum_{a=0}^\infty \bracs{\tanh \tau \cdot \tanh \tau'}^{\tr \mb{a}} \sum_{\mb{a} \geq \mb{0}: \tr \mb{a} = a}\bracs{\prod_{m=1}^M {n_m + a_m \choose a_m}}\\
&= \sum_{\mb{n} \geq \mb{0}} \bracs{\sech \tau \cdot \sech \tau'}^{\tr \mb{n} +M} \sqrt{r_\mb{n}\,s_\mb{n}} \sum_{a=0}^\infty \bracs{\tanh \tau \cdot \tanh \tau'}^{\tr \mb{a}}  {\tr \mb{n} + M -1 + a \choose a}       \label{lemmaused}\\
&= \sum_{\mb{n} \geq \mb{0}} \bracs{\sech \tau \cdot \sech \tau'}^{\tr \mb{n} +M} \sqrt{r_\mb{n}\,s_\mb{n}}  \bracs{1 -\tanh \tau \cdot \tanh \tau'}^{-\pars{\tr \mb{n} + M}}                        \label{taylorused}\\
&= \sum_{\mb{n} \geq \mb{0}}  \sqrt{r_\mb{n}\,s_\mb{n}}  \bracs{\sech\pars{\tau' - \tau}}^{\tr \mb{n} + M} \\
&= \sum_{\mb{n}\geq \mb{0}} \sqrt{r_\mb{n}\, s_\mb{n}}\,\nu^{n + M},
\end{align}
In the derivation, we have used a combinatorial lemma (See Sec.~II of the Supplementary Material of Ref.~\cite{NTG22}) to write Eq.~\eqref{lemmaused} and the Taylor expansion 
\begin{align} \label{taylor}
(1-x)^{-(n+1)} = \sum_{a=0}^\infty {n +a \choose a}\,x^a,
\end{align}
(valid for $n \geq 0$ and $\abs{x} <1$) to write Eq.~\eqref{taylorused}. Finally, Eq.~\eqref{conditionalfid} follows from using Eq.~\eqref{psileta} to set
\begin{align}
\ketas{\psi} &:= \sum_{\mb{k} \geq \mb{0}} \sqrt{p_{\mb{k}|\mb{l}}\pars{\eta}} \keta{\chi_{\mb{k} + \mb{l}}}\kets{\mb{k}},\\
\ketas{\psi'} &:= \sum_{\mb{k} \geq \mb{0}} \sqrt{p_{\mb{k}|\mb{l}}\pars{\eta'}} \keta{\chi_{\mb{k} + \mb{l}}}\kets{\mb{k}}
\end{align}
in Eq.~\eqref{states}.
\end{proof}

\end{thm}

\iffalse
\begin{figure}[htbp]
%\centering\includegraphics[trim=14mm 75mm 10mm 80mm, clip=true,width=\columnwidth]{Fig1.pdf}
%\onecolumngrid
\caption{}
\end{figure}

\begin{align}
 \left\{
	\begin{array}{ll}
		\frac{1}{\sqrt{T}} \exp\lp-i \frac{2\pi mt}{T}\rp  & \mbox{if } t \in[-T/2,T/2]  \\
		0 & \mbox{otherwise,}
	\end{array}
\right.
\end{align}
\fi

%\section{Acknowledgements}
%This work is supported by the Singapore National Research Foundation under NRF Grant No.~NRF-NRFF2011-07 and the Singapore Ministry of Education Academic Research Fund Tier 1 Project R-263-000-C06-112.

 %%%%%%%%%%%%%%%%%%%%%
\bibliography{../RNmasterbib}
%\bibliographystyle{apsrev4-1}
%\onecolumngrid
%\begin{widetext}
%\newpage
%\appendix*
%\section*{Supplementary Material}
\end{document}